\documentclass[conference,compsoc,10pt]{IEEEtran}
\pdfoutput=1
\IEEEoverridecommandlockouts
\usepackage{cite}
\usepackage{amsmath,amssymb,amsfonts}
\usepackage{algorithm}
\usepackage{algorithmic}
\usepackage{graphicx}
\usepackage{tabularx}
\usepackage{textcomp}
\usepackage{xcolor}
\usepackage{colortbl}
\usepackage{graphics}
\usepackage{subfigure}
\usepackage{makecell}
\usepackage{multirow}
\usepackage{enumitem}
\usepackage{booktabs}
\usepackage{fancyhdr}
\usepackage{balance}
\usepackage[colorlinks,
linkcolor=black,
anchorcolor=black,
urlcolor=black,
citecolor=black
]{hyperref}
\usepackage{verbatim}
\usepackage{array}
\usepackage{threeparttable}
\usepackage{bbding}
\usepackage{stfloats}

\def\BibTeX{{\rm B\kern-.05em{\sc i\kern-.025em b}\kern-.08em
    T\kern-.1667em\lower.7ex\hbox{E}\kern-.125emX}}

\makeatletter
\def\UrlAlphabet{%
      \do\a\do\b\do\c\do\d\do\e\do\f\do\g\do\h\do\i\do\j%
      \do\k\do\l\do\m\do\n\do\o\do\p\do\q\do\r\do\s\do\t%
      \do\u\do\v\do\w\do\x\do\y\do\z\do\A\do\B\do\C\do\D%
      \do\E\do\F\do\G\do\H\do\I\do\J\do\K\do\L\do\M\do\N%
      \do\O\do\P\do\Q\do\R\do\S\do\T\do\U\do\V\do\W\do\X%
      \do\Y\do\Z}
\def\UrlDigits{\do\1\do\2\do\3\do\4\do\5\do\6\do\7\do\8\do\9\do\0}
\g@addto@macro{\UrlBreaks}{\UrlOrds}
\g@addto@macro{\UrlBreaks}{\UrlAlphabet}
\g@addto@macro{\UrlBreaks}{\UrlDigits}
\makeatother

\begin{document}
	\title{Privacy-Utility Balanced Voice De-Identification Using Adversarial Examples}
    \author{
    	\IEEEauthorblockN{
        	Meng Chen\IEEEauthorrefmark{1}, 
        	Li Lu\IEEEauthorrefmark{1}, 
        	Jiadi Yu\IEEEauthorrefmark{2}, 
        	Yingying Chen\IEEEauthorrefmark{3}, 
        	Zhongjie Ba\IEEEauthorrefmark{1},
        	Feng Lin\IEEEauthorrefmark{1},
        	Kui Ren\IEEEauthorrefmark{1}
    	}
    	\IEEEauthorblockA{
    	\IEEEauthorrefmark{1}Zhejiang University
    	\IEEEauthorrefmark{2}Shanghai Jiao Tong University
    	\IEEEauthorrefmark{3}Rutgers University
    	}
    }

	\maketitle
    \thispagestyle{fancy}
    \pagestyle{fancy}
    \cfoot{\thepage}

	\begin{abstract}
        Faced with the threat of identity leakage during voice data publishing, users are engaged in a privacy-utility dilemma when enjoying convenient voice services.
        Existing studies employ direct modification or text-based re-synthesis to de-identify users' voices, but resulting in
        inconsistent audibility in the presence of human participants.
		In this paper, we propose a voice de-identification system, which uses adversarial examples to balance the privacy and utility of voice services. Instead of typical additive examples inducing perceivable distortions, we design a novel convolutional adversarial example that modulates perturbations into real-world room impulse responses. Benefit from this, our system could preserve user identity from exposure by Automatic Speaker Identification (ASI) while remaining the voice perceptual quality for non-intrusive de-identification. Moreover, our system learns a compact speaker distribution through a conditional variational auto-encoder to sample diverse target embeddings on demand. Combining diverse target generation and input-specific perturbation construction, our system enables any-to-any identify transformation for adaptive de-identification.
		Experimental results show that our system could achieve 98\% and 79\% successful de-identification on mainstream ASIs and commercial systems with an objective Mel cepstral distortion of 4.31dB and a subjective mean opinion score of 4.48.
	\end{abstract}


	\section{Introduction}
\label{sec:introduction}
Recent decades have witnessed voice input becoming one of the most prevalent methods widely deployed in various services.
Richer functional utility, including automatic speech transcription, efficient voice searching, and live language translation, thus gradually enables humans to enjoy a natural but much more intelligent experience.
However, behind the powerful utility of voice-based services, the privacy risks of voice data publishing raise extensive public concerns, especially with the wide deployment of voice-based authentication \cite{wechat,hsbc,td}. Many leading tech giants are collecting and storing users' voices in practice \cite{google,microsoft}
or even eavesdrop on users' conversations without any consent \cite{apple,alexa}. This exposes users to the risk of identity leakage by specialized Automatic Speaker Identification (ASI) tools \cite{iflyteck,azure}, or even unexpected advertisement \cite{advertising} by user profiling, malicious impersonation through voice cloning \cite{voicelone}, etc.
Caught in such a dilemma between the high utility of voice services and personal identity privacy, voice de-identification is proposed to eliminate individual traits while maintaining the linguistic content in voices for privacy-preserving voice services.
Related studies focus on the voice conversion \cite{jin2009v,abou2015a,magarinos2016p,vaidya2019y,qian2018h,zhang2020e,qian2018t,abs1905,han2020v,mohammadi2017a,srivastava2020d,turner2022g} or text-to-speech re-synthesis \cite{justin2015s,qian2018t,ahmed2020p} to transform or exclude individual features in the voice. 
However, their voice utility for human participants is significantly downgraded due to the inconsistent voiceprint and intrusive distortions, and they also exhibit insufficient adaptiveness to resist informed attacks \cite{srivastava2020e}.

Toward this end, we take a different viewpoint to balance the service utility and identity privacy for voice services, i.e., introducing adversarial examples \cite{goodfellow2014e,madry2018t,carlini2017t} as a defense tool to conceal speaker identity while remaining speech integrity and perceptual consistency. Existing adversarial example methods \cite{Yuan2018c,li2020a,chen2021w, phoneytalker,qin2019a,Schonherr2019,Abdullah2019,Wang2020} generate additive adversarial perturbations to compromise learning-based audio systems. Although they apply amplitude normalization or psychoacoustic masking to constrain the perturbation audibility, such perturbations are intrinsically additive noises, which either induce perceivable distortions \cite{Schonherr2019} or are easily filtered out \cite{hussainNDMK21,Eisenhofer2021}, making them inapplicable for voice de-identification. 
Theoretically, apart from the additive noise, the channel interference of airborne sound also includes Room Impulse Response (RIR), which can be formulated as a convolution process on the voice. RIR quantifies the over-the-air multi-path propagation of sound waves and behaves as a reverberation. Hence, it is difficult for humans to distinguish the RIR as an abnormal signal compared to the additive noises. 
Inspired by this observation, we wonder whether the RIR can be carefully crafted as a convolutional adversarial perturbation to conceal user identity while being transparent for human audibility.

Along this direction, our work aims to propose RIR-like convolutional adversarial perturbations to realize a non-intrusive and adaptive voice de-identification. To achieve this, we face several key challenges. \textit{Privacy-utility trade-off}: to balance the privacy and utility of voice services, the system needs to mislead the target ASIs but remain the linguistic content, requiring the adversarial perturbations to distort the voiceprint features without hurting the speech semantics. 
\textit{Perceivable artifact}: the perturbation injection causes perceivable artifacts and affects the perceptual quality of de-identified voices, so the convolutional adversarial perturbations need to be well crafted for non-intrusive de-identification.
\textit{Advanced attack}: to avoid the de-identified voices being re-identified and linked back to the original user, we need sufficient resistance and adaptiveness to advanced attacks, even if they are informed of our de-identification strategy already \cite{srivastava2020e}. 

In this paper, to validate the aforementioned analysis, we first conduct a preliminary study to use adversarial examples to de-identify voices. Through the experiments, we find that there is a trade-off between voice de-identification and speech utility, i.e., the adversarial example is able to de-identify voices but results in perceivable artifacts and poor adaptiveness.
To solve these problems,
we propose a non-intrusive and adaptive voice de-identification system. 
First, our system adopts a triplet loss architecture for iterative perturbation construction, whose input-specific manner enables any-to-any identity transformation so that any source user can conceal his/her identity among a large group of different target speakers for evading the detection of informed attacks. 
Then, different from the additive adversarial perturbations, our system proposes a novel convolutional adversarial perturbation method to avoid perceivable artifacts. By reshaping the convolutional adversarial perturbations into real-world RIRs, our system approximates the perturbation injection to a natural reverberation effect, remaining the perceptual consistency for human participants. 
Finally, to provide diverse targets for de-identification with limited resources, our system pre-trains a lightweight conditional variational auto-encoder at the embedding level. With this generative model, users can synthesize any target embeddings on demand, which improves the diversity of de-identified voices for adaptive de-identification. 
Experimental results show that our system could achieve effective voice de-identification on mainstream ASIs and commercial systems with satisfactory speech recognition and perceptual quality.

Our contributions are highlighted as follows:
\begin{itemize}[leftmargin=*]
	\item To the best of our knowledge, our system is the first work to employ convolutional adversarial examples to realize voice de-identification, which achieves a good balance between the privacy and utility of voice services.
	\item We propose a novel convolutional adversarial example method to modulate adversarial perturbations into real-world RIRs, which improves the perceptual consistency in terms of the voiceprint, speech content, and audio quality, realizing a non-intrusive voice de-identification.
	\item We design a triplet loss architecture for input-specific perturbation construction, and develop an embedding-level conditional variational auto-encoder to sample diverse target embeddings on demand, enabling any-to-any identity transformation for adaptive voice de-identification.
	\item Objective and subjective experiments show that our system achieves 98\% and 79\% successful de-identification on mainstream and commercial ASIs, reaching a Mel cepstral distortion of 4.31dB and a mean opinion score of 4.48.
\end{itemize}

	\section{Preliminary}
\label{sec:preliminary}
In this section, we analyze the adversarial example-based voice de-identification, and validate its feasibility to balance the privacy and utility of voice services by the preliminary experimental study.

\subsection{Background}
\label{subsec:background}

\textbf{Voice de-identification}. Driven by massive publicly-available voice data, the learning-based speech processing technology has made great strides, endowing Automatic Speaker Identification (ASI) with powerful capabilities. State-Of-The-Art (SOTA) ASIs can extract voiceprints after listening to 8$\sim$10 words, and quickly identify a single person among thousands of speakers with a low error rate \cite{voiceprint}. Although ASI is widely deployed to support service or device access security, it is also a double-edged sword that can be exploited to expose Personal Identifiable Information (PII) for identity linking or voice cloning \cite{voicelone}, raising extensive public concerns about voice data publishing. Faced with such an identity leakage threat of ASI, voice de-identification is proposed to preserve speaker privacy. Specifically, voice de-identification aims to conceal the identifiable information in the voice while remaining the linguistic texts for other downstream tasks (e.g., Automatic Speech Recognition, ASR). 
Most of existing studies follow the Voice Conversion (VC) paradigm that manipulates voiceprint through modifying acoustic features \cite{jin2009v,abou2015a,magarinos2016p,vaidya2019y,qian2018h,zhang2020e,qian2018t} or transforming speaker embeddings \cite{abs1905,han2020v,mohammadi2017a,srivastava2020d,turner2022g} to conceal the original user identity. Other studies \cite{justin2015s,qian2018t,ahmed2020p} propose to re-synthesize voices from text to completely remove individual features.
However, these researches mainly make efforts to balance the speaker identity privacy and speech content utility for machine-centric tasks, i.e., deceiving ASI and preserving ASR, rather than for human-centric experiences, i.e., the perceptual quality of de-identified voices significantly declines due to the inconsistent voiceprint and severe distortion during voice conversion and speech re-synthesis.

\textbf{Adversarial example}. Adversarial examples are originally a concept in the field of computer vision, introducing that a well-crafted noise injected into normal images can interfere with the recognition result of automatic systems, thus being a new kind of attack on automatic systems. As its development, the attack has spread to other learning-related fields, such as neural language processing, speech recognition, and speaker identification.
Most existing research optimizes and imposes \textit{Additive Adversarial Perturbations} on the original input to mislead the target systems while ensuring the imperceptibility of their perturbation injection to human beings. Following the ideas in computer vision \cite{goodfellow2014e,madry2018t,carlini2017t}, some studies \cite{Yuan2018c,li2020a,chen2021w, phoneytalker} employ amplitude normalization to constrain perturbations in the audio domain, i.e., $L_p$ normalization, but still induce perceivable distortions \cite{Schonherr2019}.
%$L_2$ normalization limits the Euclidean distance of magnitude changes and $L_\infty$ normalization restricts the maximum magnitude change.
Following studies \cite{qin2019a,Schonherr2019,Abdullah2019,Wang2020} turn to apply the Psychoacoustic Masking (PM) method to construct inaudible perturbations below the hearing threshold of the human auditory system, thus realizing highly imperceptible perturbation injection.

\subsection{System and Threat Model}
Traditional voice de-identification aims to manipulate the voiceprint for protecting user identity while remaining speech texts, but without constraints on the perceptual voiceprint consistency and audio quality. Such an approach significantly distorts the audible voices, downgrading the user experience of human listeners during the interactions. Representative interference includes the misunderstanding of human listeners during instant messaging and voice publishing on social platforms, downgrading their utility.
Toward this end, this work aims to realize a specific kind of voice de-identification, which balances the privacy (i.e., speaker identity) and utility (i.e., speech texts and user experience) for the voice services.

\begin{figure}[t]
	\centering
	\includegraphics[width=0.99\linewidth]{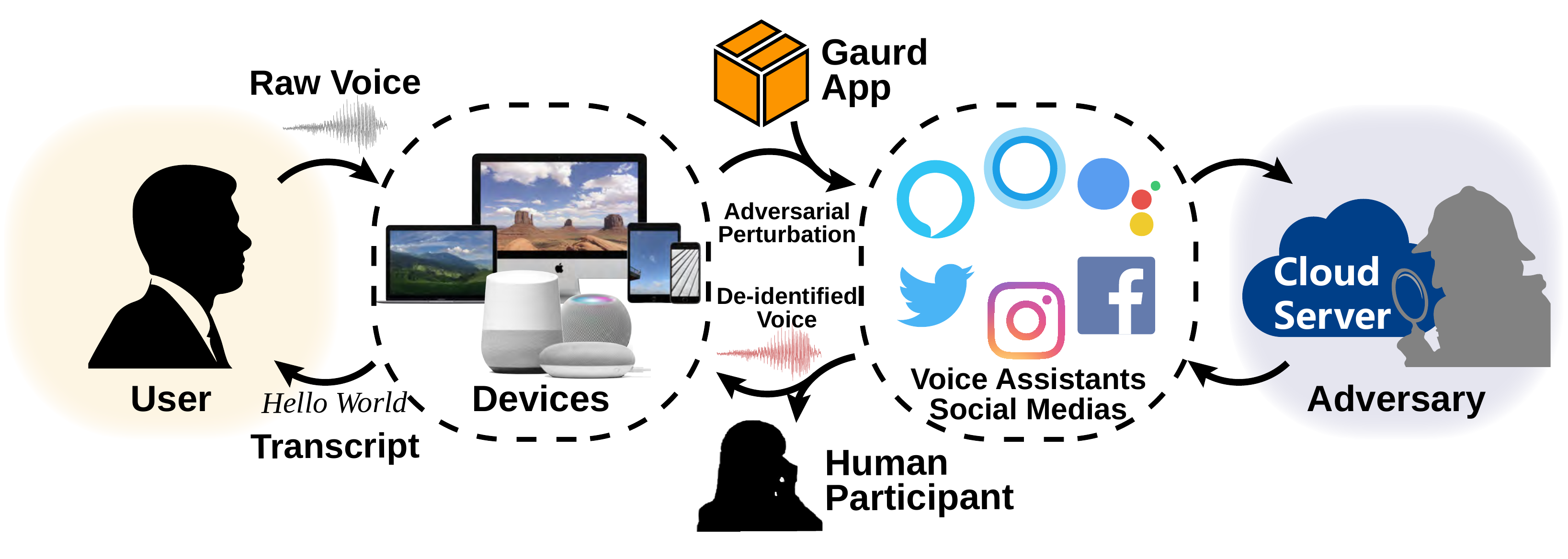}
	\caption{System and threat model.}
	\label{fig:threat-model}
\end{figure}

Figure~\ref{fig:threat-model} shows the system and threat models. In a voice service, a user's raw voice is captured by electronic devices, such as interacting with personal voice assistants, sending voice messages on instant message APPs, or sharing videos with voices on social media platforms. The captured voices are transmitted not only to human participants (e.g., social friends) but also to a cloud server for specific services (e.g., ASR). We assume the adversary can access the user's voice data on the cloud server, such as a hacker, a data analyst of a collaborated third-party company, or even the provider itself. With the collected voice data, the adversary can exploit ASI tools to extract voiceprint and explore speaker identities.
Once the user's voiceprint is exposed, it may be exploited to make user profiles for precise advertising, be shared with third parties for commercial purposes, or even be cloned to craft DeepFake voices.

To protect the user identity privacy while maintaining voice service utility, this work aims to design a guard APP at the user side for voice de-identification.
Considering the great threat to neural networks and excellent imperceptibility of adversarial examples, we intend to turn the powerful well-crafted examples into a new privacy-preserving tool to protect speaker identity from leakage to ASIs when using convenient voice services. We assume the guard APP is installed in advance and runs in the background on the user's device. Before the raw voice is uploaded to the cloud, the guard APP is invoked and imposes a subtle perturbation on the raw voice for de-identification. For convenience, the APP can be set to automation activated by microphone use events (e.g., iOS Shortcuts \cite{shortcut}, Android Anywhere \cite{anywhere}).
Such de-identified voices could conceal the user identity from the adversary's ASI while remaining perceptually consistent in both speech text and user identity for human participants.

\subsection{Feasibility Study}
\label{subsec:feasibility}
To verify our idea of leveraging adversarial examples to protect speaker privacy while maintaining user utility, we perform an experimental study to evaluate its feasibility.

\textbf{Experimental setup}.
In this experiment, we adopt an X-Vector model \cite{snyder2018x} pre-trained by Speechbrain \cite{speechbrain} as the target ASI, enrolled with 10 speakers (6 males and 4 females) from LibriSpeech corpus \cite{panayotov2015l}. Other 526 utterances from these speakers are used to generate adversarial examples with four typical additive adversarial example methods respectively, i.e., FGSM \cite{goodfellow2014e}, PGD \cite{madry2018t}, CW-$l_2$ \cite{carlini2017t}, and PM \cite{qin2019a,Schonherr2019,Abdullah2019,Wang2020}. With the goal of de-identification, we generate adversarial perturbations $\delta$ in a non-targeted manner, where the primary optimization objective is to mislead the ASI system $f(\cdot)$ to recognize the voice $x$ as any other speakers, but not the real source user $y_s$, i.e., $f(x+\delta) \ne y_s$. Specifically, FGSM and PGD employ $L_\infty$ normalization to scale perturbations within a given budget size, CW-$l_2$ introduces $L_2$ normalization in the objective to minimize the perturbation scale, and PM penalizes perturbations over the calculated hearing threshold. Finally, the generated perturbations are imposed on the original voices to derive adversarial examples. 

\begin{figure*}[t]
	\centering
	\subfigure[FGSM.]{
		\includegraphics[width=0.235\linewidth]{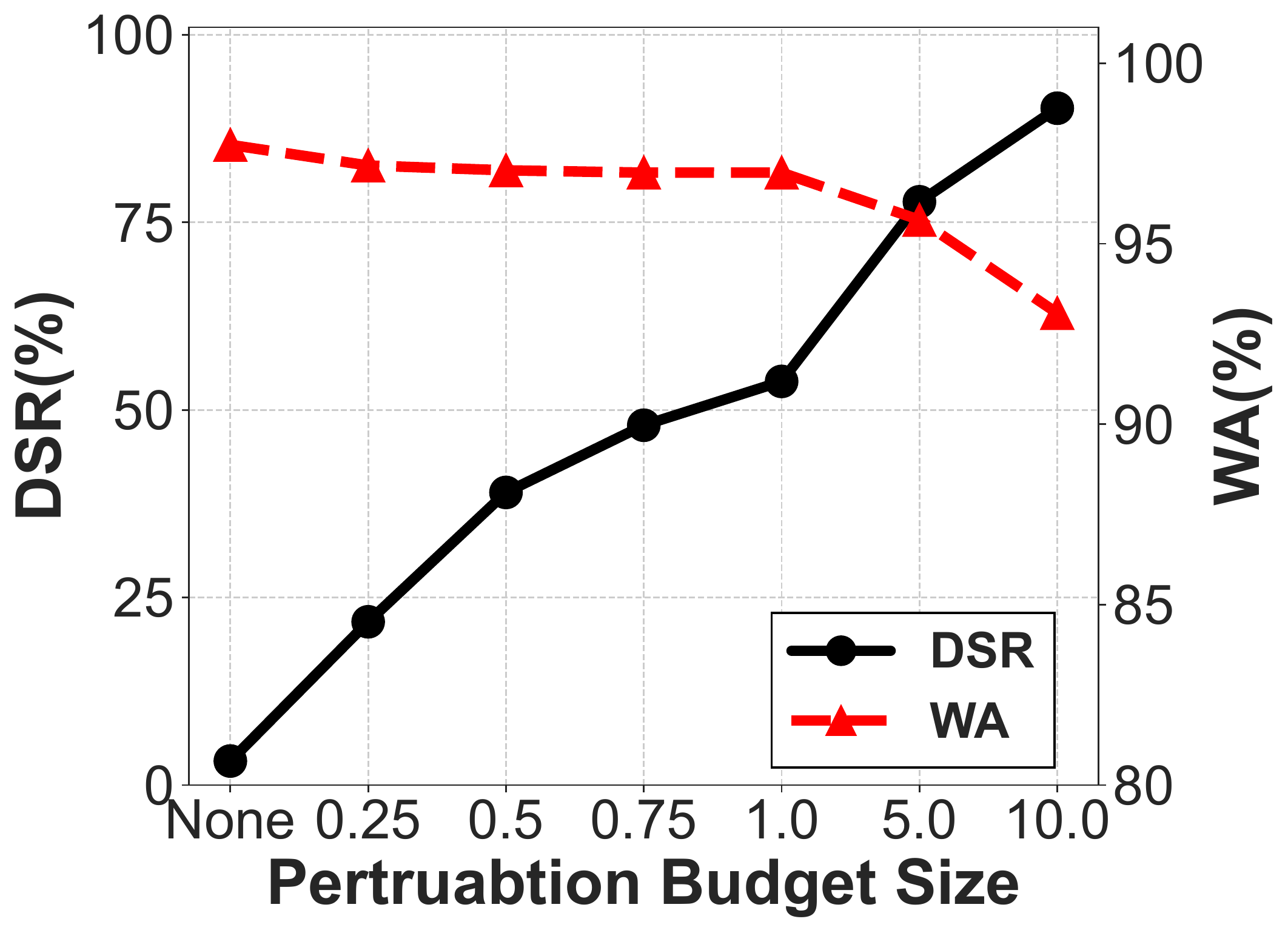}
    	\label{fig:pre-fgsm}
	}
	\hspace{-3mm}
	\subfigure[PGD.]{
		\includegraphics[width=0.235\linewidth]{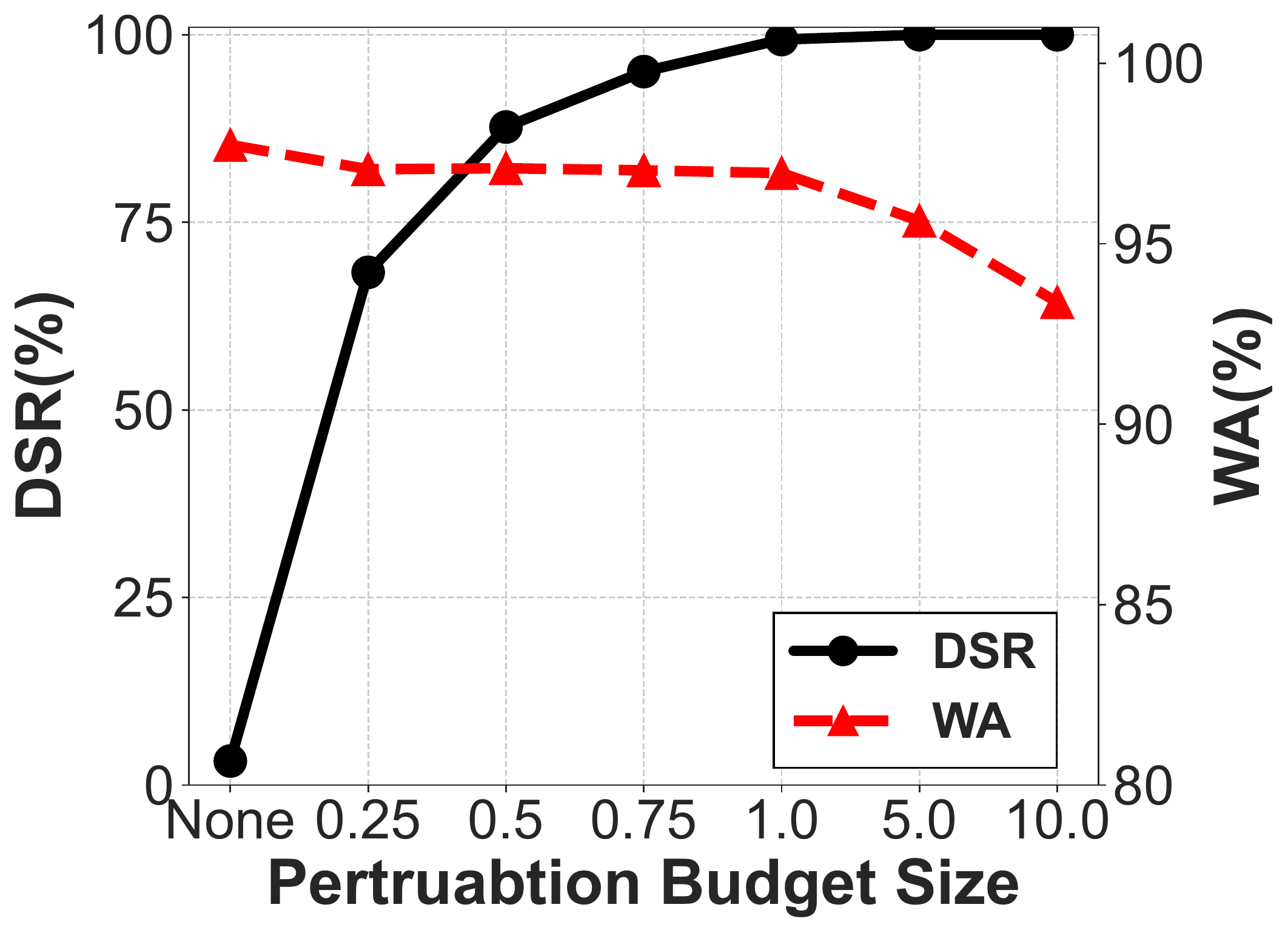}
    	\label{fig:pre-pgd}
	}
	\hspace{-3mm}
	\subfigure[CW-$l_2$.]{
		\includegraphics[width=0.235\linewidth]{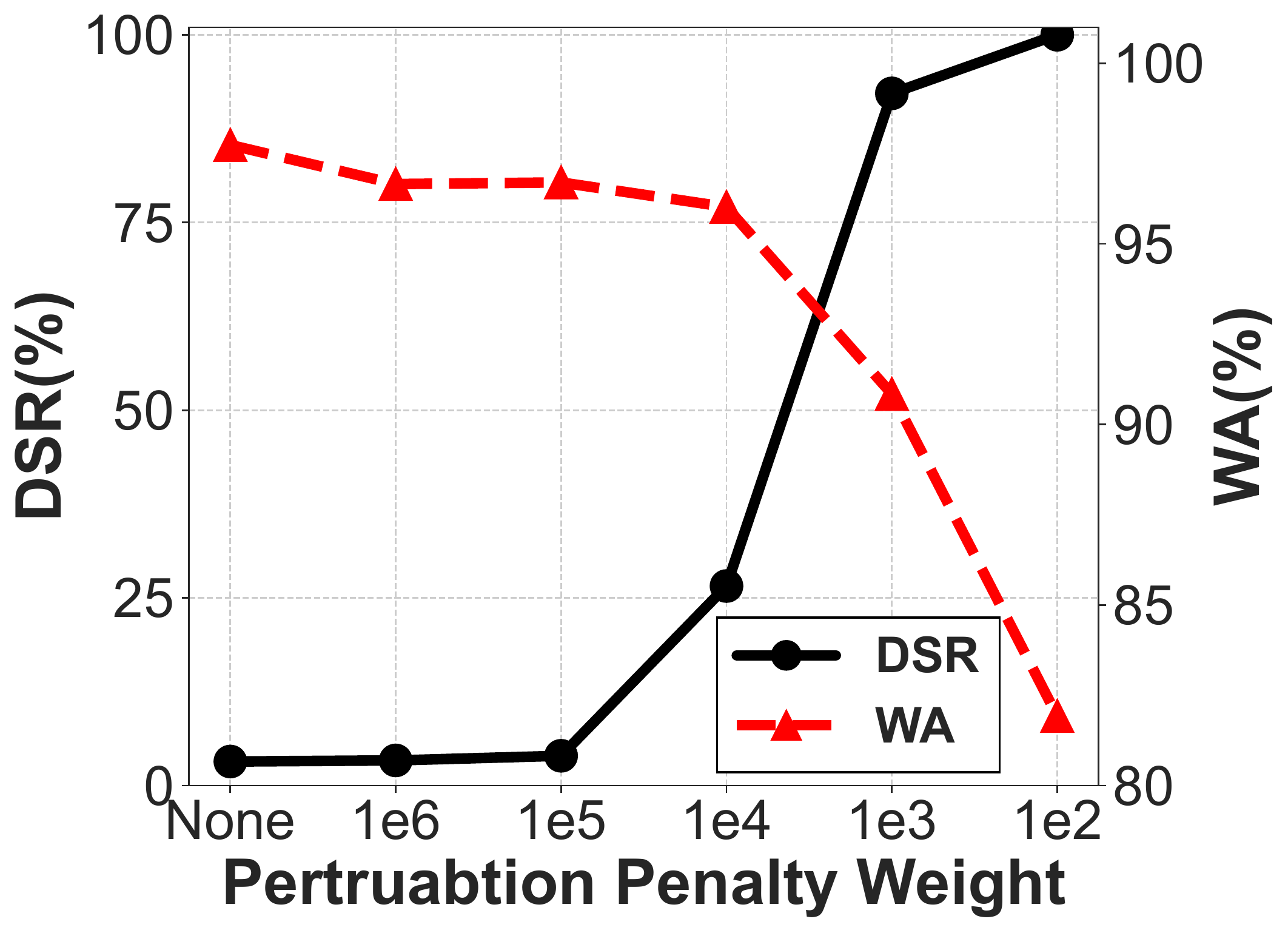}
    	\label{fig:pre-cw2}
	}
	\hspace{-3mm}
	\subfigure[PM.]{
		\includegraphics[width=0.235\linewidth]{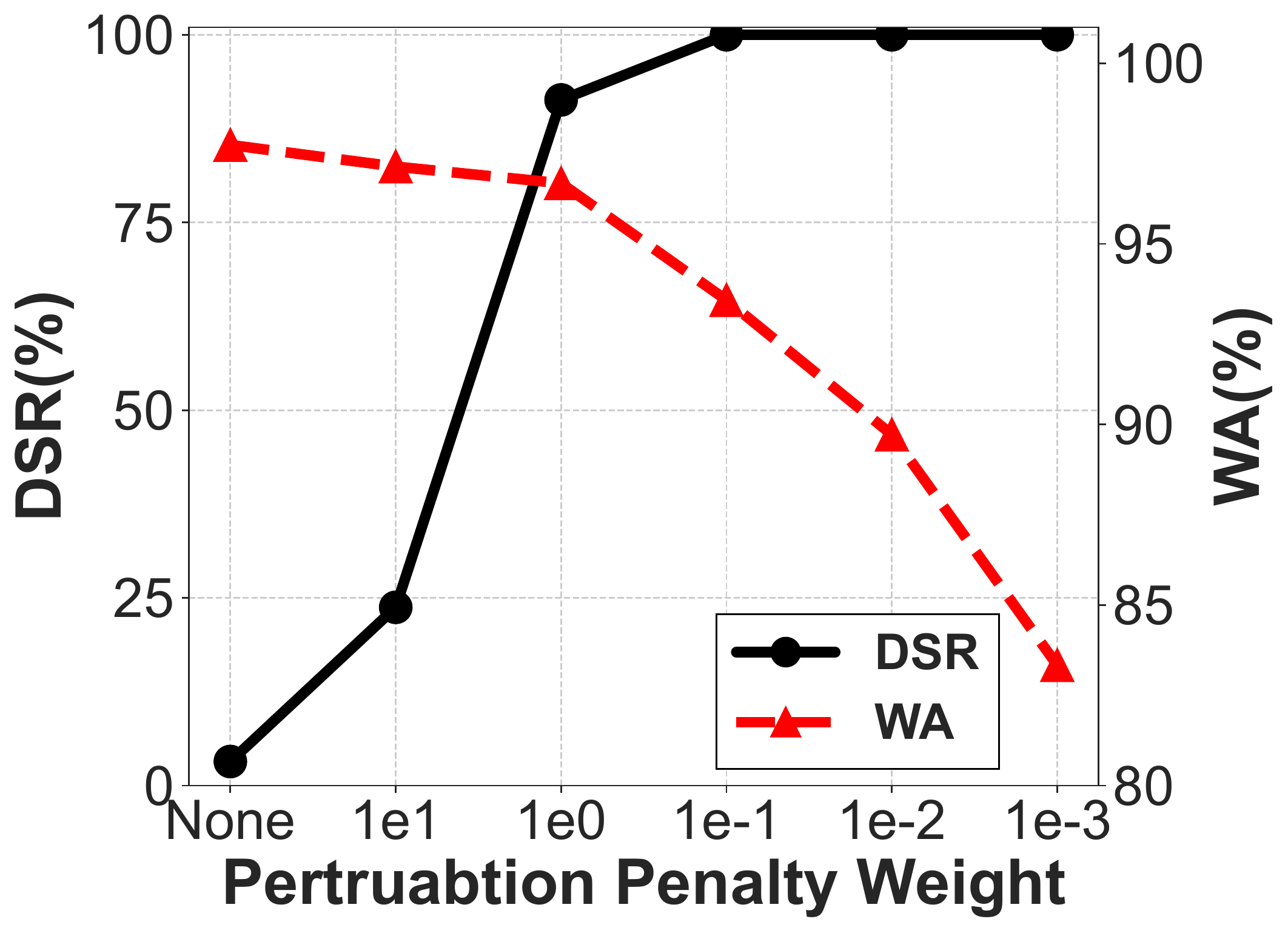}
    	\label{fig:pre-pm}
	}
	\caption{DSR and WA of different additive perturbation generation methods.}
	\label{fig:pre-dsr-wa}
\end{figure*}

\textbf{Effectiveness of voice de-identification and speech preservation}. The generated adversarial examples are fed to the ASI and an end-to-end ASR pre-trained by SpeechBrain for evaluating voice de-identification and speech preservation. We adopt the metrics including De-identification Success Rate (DSR) and Word Accuracy (WA) as defined in Section~\ref{sec:evaluation/setup}.
Figure~\ref{fig:pre-dsr-wa} shows the DSR and WA of the four additive perturbation generation methods.
As the perturbation scale grows (larger perturbation budget size for FGSM and PGD, or smaller perturbation penalty weight for CW-$l_2$ and PM), we can observe a common trend of DSR increasing and WA decreasing, which demonstrates the aforementioned privacy-utility trade-off in voice de-identification. More importantly, we find that some adversarial example methods could achieve an excellent DSR with a slight WA drop, e.g., PGD and PM approaches over 90\% DSR while sacrificing only 2\% WA. These results show the de-identification effectiveness of adversarial perturbations and their slight impact on the speech content integrity, validating the feasibility of leveraging adversarial examples for speaker identification.

\textbf{Intrusiveness to audio quality}.
To further evaluate the intrusiveness of adversarial perturbations to audio quality, we measure the signal distortions of de-identified voices. Specifically, we first select the best perturbation scale for these four methods according to Figure~\ref{fig:pre-dsr-wa}, where DSR is as high as possible while WA is at least 95\% (the parity of professional human transcriptionists). Then we calculate the Mel Cepstral Distortion (MCD) as defined in Section~\ref{sec:evaluation/setup} between the original voices and de-identified voices for each method.
As shown in Table~\ref{tab:preliminary-comparison}, compared with the DSR and WA of the baseline using original voices, PGD achieves the highest DSR of 97.54\% with a minute WA drop of 2\% but yields a high MCD of 16.68dB. With a similar WA, FGSM and CW-$l_2$ perform poorly in terms of both DSR and MCD. Thanks to the psychoacoustic masking-based design, PM reaches an excellent MCD of 3.8dB with satisfactory DSR and WA. But unfortunately, a recent work Dompteur \cite{Eisenhofer2021} has demonstrated that PM-based perturbations can be filtered out according to the same psychoacoustic principle while not hurting the normal use of audio systems. These results suggest that such additive perturbation generation schemes are either easily noticed by humans or corrupted by well-designed filters, making them insufficient for audio perturbation constraint and privacy-utility balance.

\begin{figure}[t]
	\centering
	\subfigure[Log-spectrum.]{
    	\includegraphics[height=3.1cm]{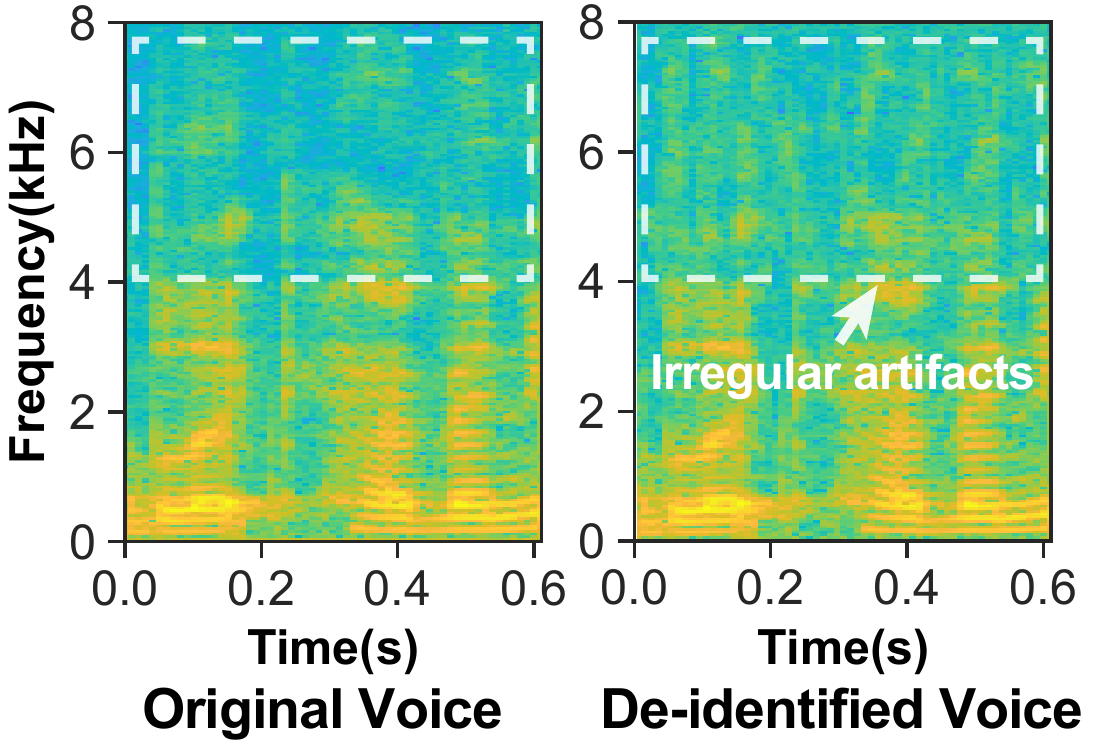}
    	\label{fig:pgd-spec}
	}
	\subfigure[T-SNE embedding.]{
		\includegraphics[height=3.1cm]{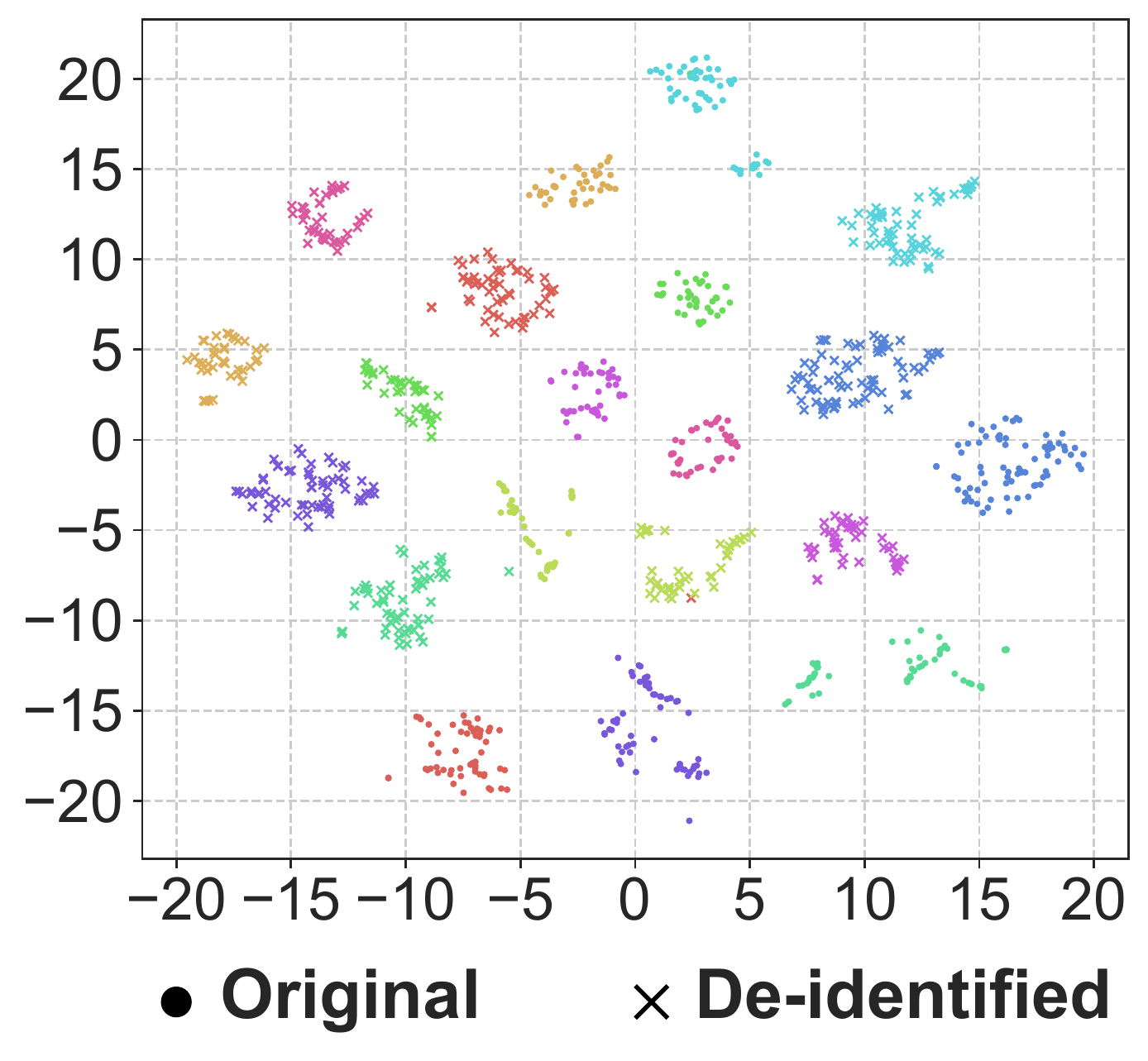}
		\label{fig:pgd-emb-tsne}
	}
	\caption{Analysis of adversarial examples generated by PGD methods.}
\end{figure}

\begin{table}[b]
    \centering
    \caption{Comparison of perturbation generation methods.}
    \label{tab:preliminary-comparison}
    \setlength\tabcolsep{5.5pt}
    % \resizebox{\linewidth}{!}{
    \begin{tabular}{ccccc}
    \toprule
        \textbf{Method} & \textbf{Constraint} & \textbf{DSR(\%)} & \textbf{WA(\%)} & \textbf{MCD(dB)} \\
        \midrule
        Baseline & None & 3.21 & 97.74 & None \\
        FGSM & $L_\infty$ Normalization & 77.73 & 95.66 & 8.28$\pm$1.67 \\
        PGD & $L_\infty$ Normalization  & \textbf{97.54} & 95.73 & 16.68$\pm$3.56 \\
        CW-$l_2$ & $L_2$ Normalization       & 26.57 & 96.02 & 6.27$\pm$1.64 \\
        PM & Psyacoustic Masking    & 91.32 & \textbf{96.67} & \textbf{3.80$\pm$0.91} \\
        \bottomrule
    \end{tabular}%}
\end{table}

\textbf{Analysis of generated adversarial examples}. To further investigate the signal distortion induced by additive perturbations, we study the original voice and adversarial example generated by PGD in both time and frequency domains. As the spectrum shown in Figure~\ref{fig:pgd-spec}, we can observe obvious irregular artifacts introduced by additive adversarial perturbations with high frequency and energy, which can be easily perceived by human ears.
Additionally, we apply t-SNE to visualize the speaker embeddings extracted from original voices and adversarial examples. As shown in Figure~\ref{fig:pgd-emb-tsne}, for each speaker, the embeddings of adversarial examples shift away from those of original voices, thus successfully misleading the ASI to identify them as different individuals.
However, all the embeddings of adversarial examples from each speaker still cluster together with low diversity, which exposes speakers to the risk of re-identification, leading to poor adaptiveness to informed service providers. We find this problem is attributed to the non-targeted perturbation optimization that always searches for similar gradient directions to update perturbations for each utterance.
These observations inspire us to generate more diverse and audibility-friendly adversarial examples for protecting user identity against adaptive adversaries while maintaining user experience in voice services.

	\section{System Design}
\label{sec:system-design}
In this section, we illustrate the design details of our system for preserving the speaker identity privacy and maintaining the voice service utility.

\subsection{Overview}
To balance the privacy and utility of voice services, we propose a non-intrusive and adaptive speaker de-identification system using adversarial examples. Instead of directly manipulating the voiceprint, the basic idea is to modulate \textit{Convolutional Adversarial Perturbations} into the real-world room impulse responses to produce adversarial examples, which approximates the perturbation inject as a natural reverberation effect to minimize perceivable artifacts and ensure the perceptual consistency. Moreover, our system pre-trains a lightweight \textit{Conditional Variational Auto-Encoder} to generate diverse targets on demand and then constructs adversarial examples in an input-specific manner. This enables any source user to disguise a large group of target speakers in different utterances, thus improving the diversity and adaptiveness of de-identified voices.

\begin{figure}[t]
	\centering
	\includegraphics[width=0.9\linewidth]{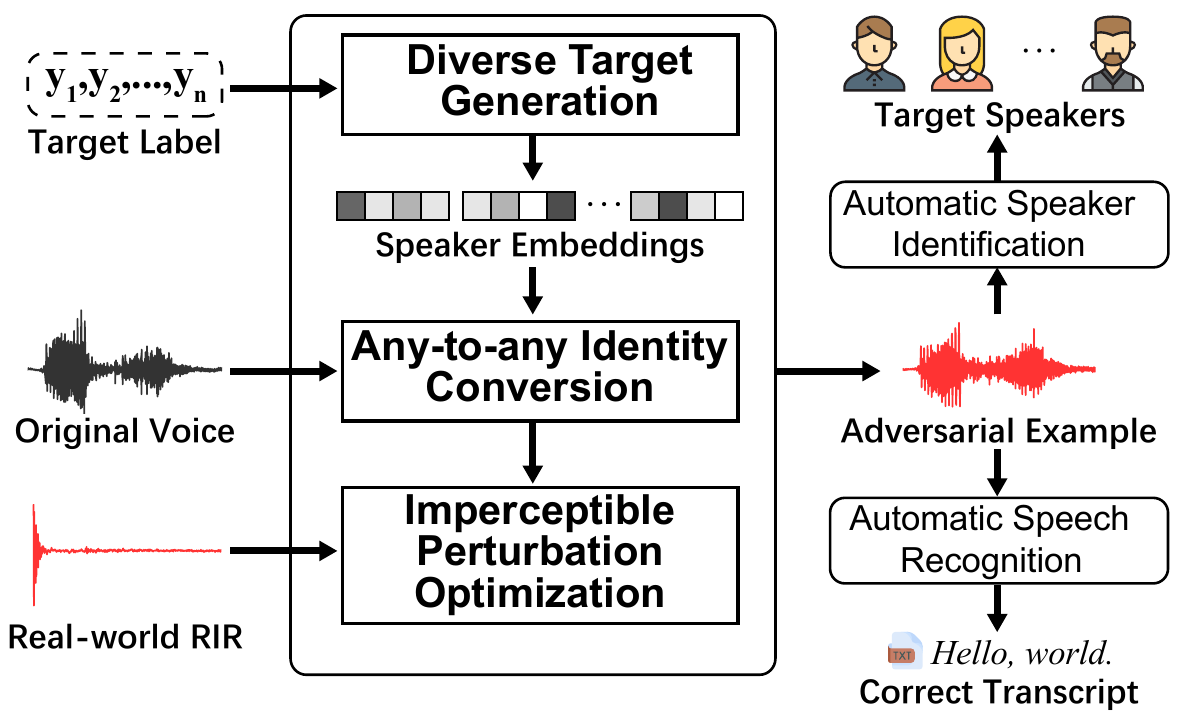}
	\caption{System overview.}
	\label{fig:system-overview}
\end{figure}

Figure \ref{fig:system-overview} shows the system overview. In \textit{Any-to-any identity conversion}, we first adopt a triplet loss architecture to construct input-specific adversarial examples, which converts the identity of any source user to any desired target speaker. 
To minimize the perceivable distortions and artifacts resulting from typical additive perturbation injection, we propose a novel convolutional adversarial perturbation method in \textit{Imperceptible perturbation optimization}. Such convolutional adversarial perturbations are reshaped into real-world room impulse responses and then modulated into the natural reverberation effect of airborne sound propagation, thus remaining perceptually consistent voiceprints and high speech quality for human participants.
In \textit{Diverse target generation}, we design a lightweight conditional variational auto-encoder at the embedding level, instead of storing pre-collected data as a pool, to synthesize samples of target speakers on demand for reducing the resource demand on users' devices.
By explicitly modulating prior identity knowledge to the generative model, our system could learn a compact speaker identity distribution and sample the embeddings of any target, even a pseudo speaker as instances for perturbation optimization, which improves the diversity and adaptiveness of de-identified voices.
Consequently, these adversarial examples would be identified as different target speakers instead of the source user by ASI, while preserving the correct transcription from ASR. 

\subsection{Any-to-any Identity Conversion}
\label{subsec:perturb-construct}
To protect the user identity from disclosure, we first construct adversarial perturbations to impose on the user's raw voice for de-identification.

Typically, there are two different manners of adversarial perturbation construction, i.e., the non-targeted and targeted manners. As demonstrated in Section~\ref{subsec:feasibility}, the non-targeted manner yields highly-similar adversarial examples that may be re-identified by the voice service provider. Hence, we turn to consider the targeted manner, which shifts the source identity feature towards a given target, thereby enabling the user to have different identities with different targets. Since we have no prior knowledge about the enroll set of adversary's ASI, we can not guarantee full de-identification using only targeted adversarial examples, which may still be identified as the source user rather than other enrolled speakers. Therefore, we introduce a triplet loss architecture \cite{schultzJ2003l} to combine the non-targeted and targeted manners for effective voice de-identification.

The core idea of triplet loss is to quantify the distance among samples in the latent space, and project the anchor and positive samples in the nearby region, whereas the anchor and negative samples are far away from each other. Specifically, we define the speaker embeddings of the source voice and target voice (i.e., $f(\mathbf{x_s})$ and $f(\mathbf{x_t})$) as the negative and positive samples respectively, and regard the speaker embedding of adversarial example (i.e., $f(\mathbf{x_s'})$) as the anchor sample, where $f(\cdot)$ denotes the ASI model for embedding extraction.
Then, we build a triplet and derive the triplet loss:
\begin{equation}
	\label{equ:weighted-triplet-loss}
	\mathcal{L}(x_s') = D(f(\mathbf{x_s'}),f(\mathbf{x_t}))-D(f(\mathbf{x_s'}),f(\mathbf{x_s})),
\end{equation}
where $D(\cdot)$ refers to the distance metric of speaker embeddings (e.g., PLDA \cite{dehak2010front}, cosine distance).
With the objective in Equation~(\ref{equ:weighted-triplet-loss}), we iteratively optimize the adversarial example in a gradient descent way under the explicit guidance of the triplet, enabling the user's identity to be close to another target speaker while far away from the source user to the greatest extent.

Since the adversarial example is constructed from specific source and target voices as needed, such an input-specific manner empowers our system any-to-any voice de-identification.
On the one hand, by selecting different targets from a pool of pre-collected voices to build triplets, this manner allows a source user to disguise a group of different speakers among different voices, further increasing the difficulty of identity detection by ASIs. On the other hand, any source users could directly deploy and apply our system without any additional enrollment, enabling a user-friendly experience.

\subsection{Imperceptible Perturbation Optimization}
\label{subsec:perturb-optimize}
Compared with voice conversion and speech re-synthesis that directly modify voiceprint or generate audio with significant distortions, we impose subtle adversarial perturbations on raw voices, which remains a perceptually consistent voiceprint for human participants while effectively deceiving ASIs. 
However, as mentioned in the preliminary study in Section~\ref{subsec:feasibility}, typical additive adversarial perturbations either still produce perceivable artifacts or are easily filtered out.
Instead, we propose a novel imperceptible perturbation optimization approach to minimize the intrusiveness to human participants.

\begin{figure}[t]
	\centering
	\subfigure[RIR illustration.]{
		\includegraphics[height=2.4cm]{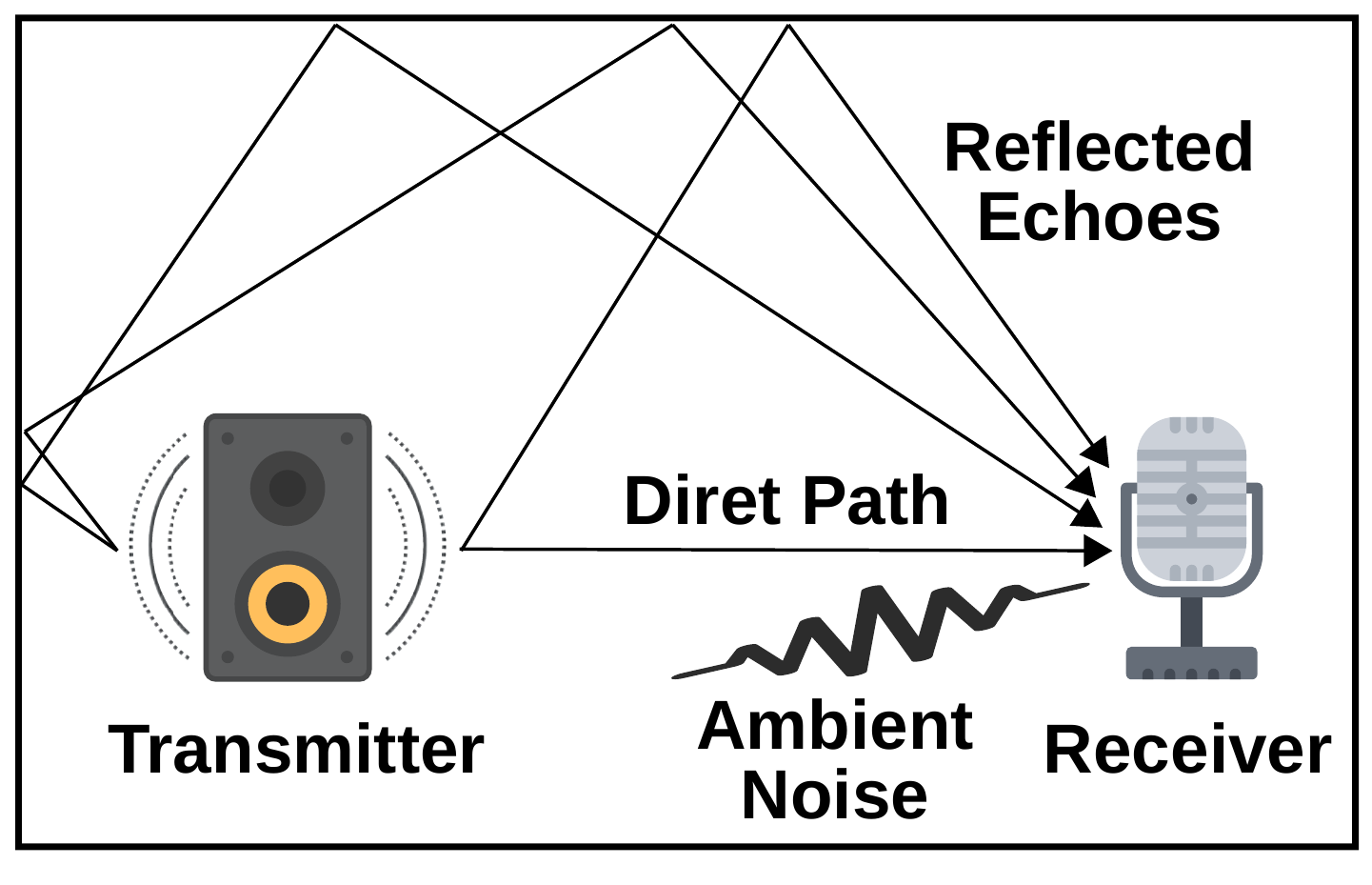}
		\label{fig:rir-illustrate}
	}
	\hspace{-2mm}
	\subfigure[RIR example.]{
		\includegraphics[height=2.4cm]{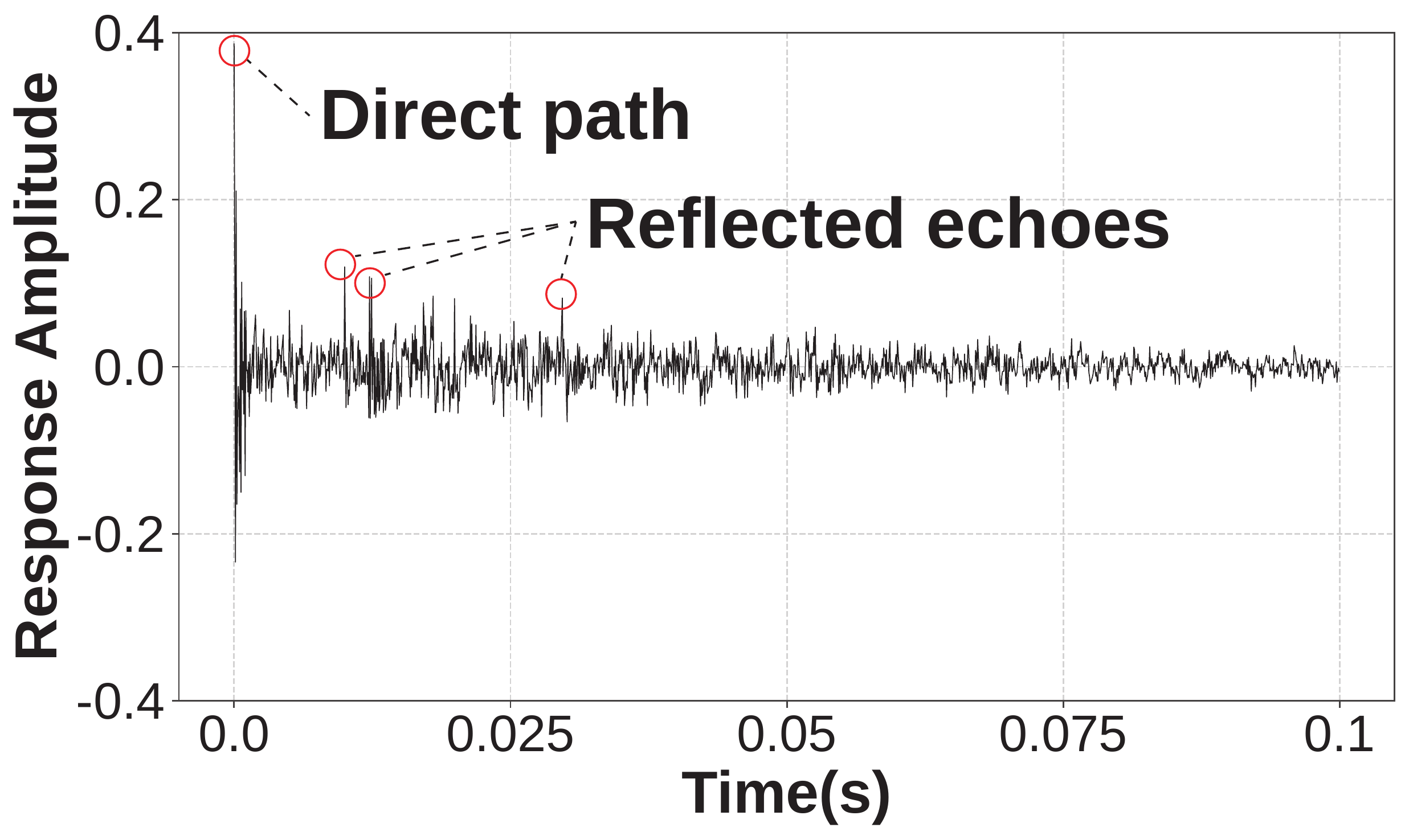}
		\label{fig:rir-example}
	}
	\caption{Illuatration and example of real-world RIR.}
	\label{fig:rir}
\end{figure}

\begin{figure*}[t]
	\centering
	\includegraphics[width=0.88\linewidth]{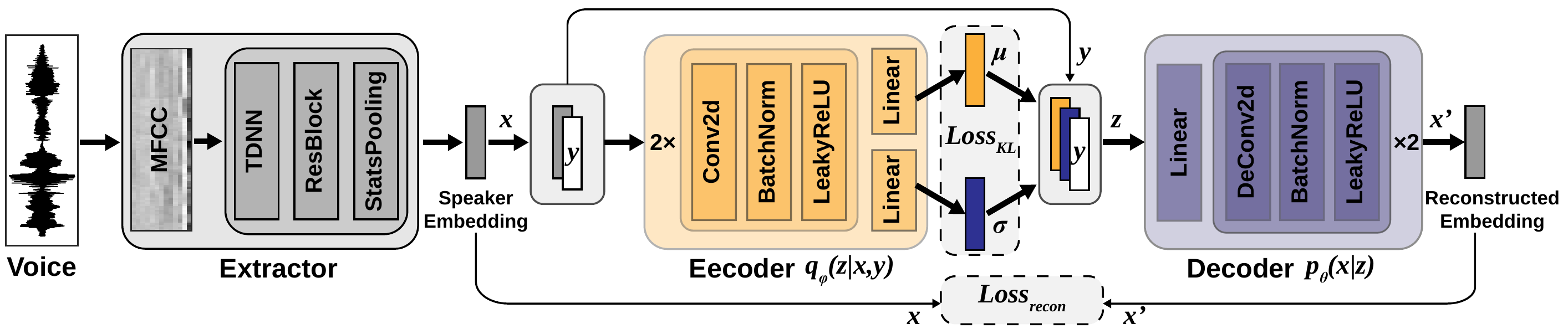}
	\caption{Network architecture of the embedding-level $\beta$-CVAE.}
	\label{fig:beta-cvae}
\end{figure*}

Theoretically, the over-the-air propagation of sound waves involves two kinds of channel interference, i.e., additive noise and Room Impulse Response (RIR). Different from the typical adversarial example methods that impose perceivable additive noises on original voices, we turn to explore the RIR. Theoretically, RIR quantifies the multi-path effect during the sound propagation in physical space. As shown in Figure~\ref{fig:rir}, the sound waves coming from the transmitter propagate omnidirectionally in the room, so the received signal at the receiver mainly includes: (1) the waves travel through the direct path; (2) the echoes reflected from surrounding walls with different delays; (3) the ambient noise. Such a roughly linear time-invariant process can be formulated as a convolution operation on the original voice, which behaves as a reverberation effect and is difficult to be distinguished as an anomalous signal by humans. Inspired by this observation, we propose to conduct RIR-like convolutional adversarial perturbations for realizing imperceptibility.

Specifically, unlike the additive perturbation scheme, i.e, $\mathbf{x_s'}=\mathbf{x_s}+\delta$, we convolve the perturbation and original voice to produce adversarial examples, i.e., $\mathbf{x_s'}=\mathbf{x_s}\ast\mathbf{\delta}$. According to the property of the convolution operation, we have:
\begin{equation}
    \label{equ:convolution}
    \text{FFT}(\mathbf{x_s}\ast\mathbf{\delta})=\text{FFT}(\mathbf{x_s})\times\text{FFT}(\mathbf{\delta}).
\end{equation}
Hence, the convolution in the time domain is equivalent to multiplication in the frequency domain. In other words, our convolutional perturbation is essentially a filter, which determines the significance of different frequency components in the original voice. This filter can serve as a promising approach to manipulating acoustic features in voices for de-identification, such as pitch and harmonics.
Then, we introduce the real-world RIR as a reference to guide the optimization of convolutional adversarial perturbations. Specifically, after expanding Equation~(\ref{equ:weighted-triplet-loss}), we penalize the convolutional perturbation change in the following optimization objective:
\begin{equation}
	\label{equ:rir-loss}
	\begin{split}
	\mathcal{L}(\mathbf{\delta'}) & = D(f(\mathbf{x_s}\ast\mathbf{\delta'}),f(\mathbf{x_t})) \\
	& - D(f(\mathbf{x_s}\ast\mathbf{\delta'}),f(\mathbf{x_s}))+\alpha \Vert \mathbf{\delta'}-\mathbf{\delta} \Vert_p,
	\end{split}
\end{equation}
where $\mathbf{\delta}$ is the real-world RIR, $\mathbf{\delta'}$ is the convolutional adversarial perturbation to be optimized, and $\alpha$ is the penalty weight. In our implementation, $L_2$ normalization is adopted according to empirical studies. By optimizing this objective, we reshape the convolutional adversarial perturbations as real-world RIRs to approximate a natural reverberation effect.

Benefit from such RIR-like convolutional perturbations, each sampling point of our adversarial example is derived from the weighted combination of the past sampling points of the original voice, leading to less artifact and distortion. Based on this, we can achieve effective and imperceptible voice de-identification, providing perceptually consistent voiceprint and audio quality for human participants.

\subsection{Diverse Target Generation}
\label{sec:target_gen}

Using the constructed convolutional perturbations, our system can mislead ASIs to regard the source user as a variety of different target speakers, preserving the user identity privacy.
To realize a sufficiently diverse de-identification, previous work collects and stores a large number of target speaker voices, whose speaker embeddings are also extracted and updated frequently. 
Considering the unacceptable memory occupancy and computing overload of voice collection and update, we propose to pre-train a lightweight generative network for expressive and diverse target sample generation.

In order to generate samples with desired speaker identity as targets, we employ a Conditional Variational Auto-Encoder (CVAE), for its strong ability to model continuous distribution and learn semantic representation. Since the adversarial perturbation construction only involves the triplet distance of speaker embeddings in the latent space, it introduces redundant voice-embedding transformations to directly build CVAE at the voice sample level. To this end, we design a lightweight embedding-level $\beta$-CVAE, serving as a target speaker embedding generator. As shown in Figure \ref{fig:beta-cvae}, we first extract speaker embeddings from a multi-speaker corpus with an extractor. Following the framework of mainstream ASIs \cite{snyder2018x,desplanques2020e}, the extractor converts the voice in the corpus to the frequency domain by short-time Fourier transform and then extracts Mel Frequency Cepstral Coefficients (MFCC) \cite{fayek2016s} as acoustic features. The acoustic features are fed to a Time Delay Neural Network (TDNN) to build the frame-level temporal context, and then input to residual blocks to model global channel interdependences. Finally, statistics pooling is performed to aggregate utterance-level features as the output speaker embedding. This embedding extraction process serves as a data preparation, where we directly employ mainstream open-source ASIs as the extractor to generate a large number of speaker embeddings to construct an embedding dataset for the following $\beta$-CVAE training.

Taking the extracted speaker embedding $\mathbf{x}$ as input, the $\beta$-CVAE generates a new embedding $\mathbf{x'}$ through an encoder-decoder architecture, where the encoder $q_{\phi}$ squeezes the embedding $\mathbf{x}$ into a latent variable $\mathbf{z}$, and the decoder $p_{\theta}$ reconstructs a new embedding $\mathbf{x'}$ through deep neural networks. To learn the underlying identity semantics, the speaker embedding $\mathbf{x}$ is concatenated with the one-hot embedding of the corresponding identity label $\mathbf{y_t}$ and then fed to the encoder. The encoder stacks multiple down-sample blocks consisting of convolution layers and batch normalization with LeakyReLU activation. According to the variational interference, the encoder output $\mathbf{z}$ is assumed to follow a Gaussian distribution: $q_{\phi}(\mathbf{z}|\mathbf{x},\mathbf{y})\sim\mathcal{N}(\mathbf{\mu},\mathbf{\sigma}^2\mathbf{I})$, where the mean vector $\mathbf{\mu}$ and the diagonal covariance $\mathbf{\sigma}^2\mathbf{I}$ are derived from the following two parallel linear layers. After that, the latent variable $\mathbf{z}$ can be sampled from the distribution, which is reparameterized using the reparameterization tricks \cite{kingmaW2013a} in practice: $\mathbf{z}=\mathbf{\mu}+\mathbf{\sigma}\odot\epsilon$, where $\epsilon\sim\mathcal{N}(\mathbf{0},\mathbf{I})$. Similar to the encoder input, we also explicitly modulate the prior $\mathbf{y_t}$ into the latent variable $\mathbf{z}$ as the input of the decoder. The decoder reforms the latent vector by a linear layer, and then reconstructs a new embedding $\mathbf{x'}$ with multiple up-sample blocks, each of which consists of deconvolution layers and batch normalization with LeakyReLU activation. Finally, the entire generative model can be trained with:
\begin{equation}
	\label{equ:beta-cvae-loss}
	\begin{split}
		\mathcal{L}_{\beta-CVAE} &= \mathbb{E}[\Vert \mathbf{x}-\mathbf{x'} \Vert_2] \\
		& \quad +\beta KL(\mathcal{N}(\mathbf{z}|\mathbf{\mu},\mathbf{\sigma}^2\mathbf{I}) \Vert \mathcal{N}(\mathbf{z}|\mathbf{0},\mathbf{I})),
	\end{split}
\end{equation}
where the first term is the reconstruction error, restricting the reconstruction bias of the output embedding, and the second term is the KL divergence regularization, forcing the latent space to approximate a continuous zero-mean unit-variance Gaussian. And the weight $\beta$ is used to balance the reconstruction quality and the latent space continuity.

With the well-trained $\beta$-CVAE, our system could generate diverse target speaker embeddings for adversarial perturbation construction. Specifically, given the desired identity label, the target embedding can be derived from: (1) \textit{reconstruction}: the generator encodes the speaker embedding and then decodes the latent variable to a new embedding with the corresponding identity. (2) \textit{sampling}: the generator directly samples a stochastic variable in the latent space and synthesizes a new embedding only with the decoder. (3) \textit{interpolation}: the generator performs semantic interpolation between latent variables of different speakers to synthesize new embeddings of unreal speakers, achieving an effect of "walking in the latent space" \cite{hsu2017l}. As a result, our system could sample diverse speaker embeddings with the desired identity from a learned Gaussian distribution without any voice input, or even create unreal embeddings as "pseudo speaker" for improving the identity diversity. Moreover, only the pre-trained decoder module is needed for actual deployment, thus significantly reducing the demand for computing and storage resources.

Finally, we summarize our complete adversarial perturbation construction process for voice de-identification in Algorithm~\ref{alg:any-to-any-algorithm}.

\begin{algorithm}[t]
	\caption{\label{alg:any-to-any-algorithm}Adversarial Perturbation Construction}
	\begin{algorithmic}[1]
		\REQUIRE ASI model $f(\cdot)$ with distance metric $D(\cdot)$ \\
		\quad\quad Voice of source speaker $\mathbf{x_s}$ \\
		\quad\quad Label of target speaker $\mathbf{y_t}$ \\
		\quad\quad Real-world RIR $\mathbf{\delta}$ \\
		\quad\quad Pre-trained $\beta$-CVAE dectoder $d(\cdot)$\\
		\quad\quad Perturbation penalty weight $\alpha$ \\
		\quad\quad Learning rate $\eta$ \\
		\ENSURE Well-crafted convolutional perturbation $\mathbf{\delta'}$
		\STATE $\mathbf{\delta'}\gets\mathbf{\delta}$ \qquad $\triangleleft$ Perturbation initialization
		\STATE $\mathbf{p} \gets d(\mathbf{y_t})$ \qquad $\triangleleft$ Target generation
		\STATE $\mathbf{n} \gets f(\mathbf{x_s})$
		\FOR {each iteration}
		\STATE $\mathbf{a} \gets f(\mathbf{x_s}\ast\mathbf{\delta')}$  \qquad $\triangleleft$ Convolution Operation
		\STATE $\mathcal{L}(\mathbf{a},\mathbf{p},\mathbf{n}) \gets D(\mathbf{a},\mathbf{p})-D(\mathbf{a},\mathbf{n})+\alpha\Vert\mathbf{\delta'}-\mathbf{\delta}\Vert_2$
		\STATE $\mathbf{\delta'}\gets\mathbf{\delta'}-\eta\nabla_\mathbf{\delta'}\mathcal{L}(\mathbf{a},\mathbf{p},\mathbf{n})$
		\ENDFOR
	\end{algorithmic}
\end{algorithm}

	\section{Evaluation}
\label{sec:evaluation}
In this section, we evaluate the performance of our system on open-source ASI systems as well as commercial platforms with a multi-speaker voice corpus.
\subsection{Experimental Setup}\label{sec:evaluation/setup}
\textbf{Voice dataset}. We evaluate our system with a large-scale multi-speaker corpus \textit{LibriSpeech} \cite{panayotov2015l}, which contains over 100 hours of English voices from 281 speakers, covering a wide range of different accents, professions, and ages. In this corpus, the voices vary from several seconds to tens of seconds, under a sampling rate of 16kHz.
In our experiments, we enroll the 40 speakers (20 males and 20 females) from the test-clean subset in the ASI systems as users. And the remaining 2,189 utterances are used to test the ASI systems and construct adversarial examples. In addition, the voices from the remaining 241 speakers are used to pre-train the embedding-level $\beta$-CVAE for target speaker generation. Note that the 40 source users and 241 target speakers are disjoint, i.e., the actual users are unseen for our $\beta$-CVAE training.

\textbf{ASI systems}. We implement multiple State-Of-The-Art (SOTA) speaker identification models as target ASI systems, including D-Vector \cite{variani2014d}, DeepSpeaker \cite{li2017d}, X-Vector \cite{snyder2018x},
and Ecapa-TDNN \cite{desplanques2020e}. These ASIs have different feature extractors, network architectures and parameter settings. We train D-Vector and DeepSpeaker by ourselves for the lack of available open-source pre-trained models, while X-Vector and Ecapa-TDNN are pre-trained by SpeechBrain \cite{speechbrain}, available on Hugging Face \cite{hugginface}.

\textbf{Implementation details}. For the $\beta$-CVAE training, we first extract speaker embeddings from the dataset of 241 speakers. Then we concatenate the extracted embeddings with the one-hot embedding of identity labels as the encoder input and derive the synthesized embeddings.
We set $\beta$ to 2 in Equation~(\ref{equ:beta-cvae-loss}), and pre-train the $\beta$-CVAE with an Adam optimizer (momentum=(0.9, 0.999), learning rate=0.001) for 30 epochs to derive a compact distribution.
After the training, we only remain the pre-trained decoder to sample target embeddings and build triplets, as mentioned in Section~\ref{sec:target_gen}. Then, we construct convolutional adversarial perturbations by optimizing the triplet loss with perturbation penalty in Equation~(\ref{equ:rir-loss}), where $\alpha$=5000 by default. We optimize the perturbations until 200 iterations or early stopping with a patient of 10. The well-crafted perturbations are normalized to the identical power and then convolved with the user's voices to produce adversarial examples for voice de-identification.

\begin{figure}[t]
    \centering
    \includegraphics[width=0.8\linewidth]{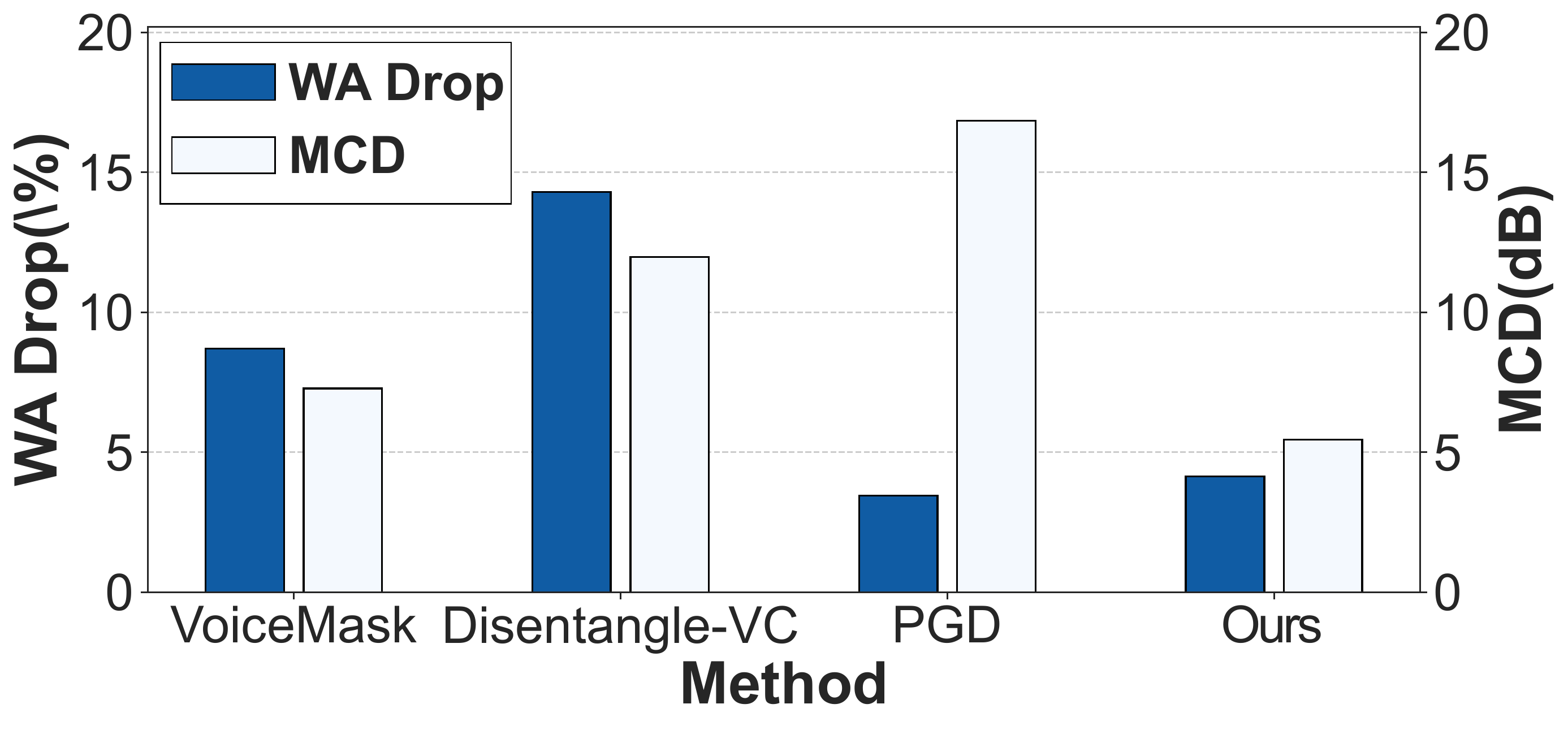}
    \caption{WA drop and MCD of our system and other three methods.}
    \label{fig:sota-comparison}
\end{figure}

\textbf{Evaluation metrics}.
(1) \textit{De-identification Success Rate (DSR)}: $\text{DSR}=\frac{X}{Y}$, where $Y$ is the total number of user voices and $X$ is the number of successful de-identification. A higher DSR means more effective de-identification.
(2) \textit{Word Accuracy (WA)}: $\text{WA}=\frac{C-I}{N}$, where $N$ is the sentence length, $C$ and $I$ refer to the number of correct words and extra inserted words in the transcript respectively. A higher WA from the same ASR system indicates more complete linguistic content.
(3) \textit{Mel Spectral Distortion (MCD)}: $\text{MCD}=10\ln{10}\sqrt{2\sum_{i=1}^{24}{\Vert mc_r(i)-mc_t(i)\Vert_2}}$, an objective audio distortion measurement that quantifies the distance between the MFCCs of the reference and target voices ($mc_r, mc_t$). Typically, an MCD below 8dB is acceptable for an ASR system while between 4.5dB$\sim$6.0dB is needed for voice conversion systems \cite{cmu_speech}.
(4) \textit{Mean Opinion Score (MOS)}: a numerical measure of the human-judged quality of a voice, where the ratings are mapped to 5 levels: excellent(5), good(4), fair(3), poor(2) and bad(1).
(5) \textit{Real-Time Ratio (RTR)}: $\text{RTR}=\frac{T_c}{T_d}$, where $T_c$ and $T_d$ are the time cost and voice duration respectively. An RTR below 1 is considered real-time.

\begin{table}[b]
	\centering
	\caption{Overall performance of on SOTA ASIs.}
	\label{tab:overall}
	\setlength\tabcolsep{4pt}
	\begin{tabular}{c|cc|ccc}
		\toprule
		\multirow{2}{*}{\textbf{ASI}}
		& \multicolumn{2}{c|}{\textbf{Original}} & \multicolumn{3}{c}{\textbf{De-identified}} \\
		& \textbf{DSR(\%)} & \textbf{WA(\%)} & \textbf{DSR(\%)} & \textbf{WA(\%)} & \textbf{MCD(dB)} \\
		\midrule
		D-Vector & 8.05 & 97.23 & 99.68 & 96.65 & 3.83$\pm$0.74 \\
		DeepSpeaker & 7.11 & 97.23 & 100.00 & 94.02 & 3.70$\pm$0.51 \\
		X-Vector & 5.42 & 97.23 & 98.51 & 93.13 & 5.45$\pm$0.82 \\
		Ecapa-TDNN & 1.25 & 97.23 & 99.29 & 94.11 & 4.26$\pm$0.54\\
		\bottomrule
	\end{tabular}
\end{table}

\begin{figure}[t]
    \centering
    \includegraphics[width=0.8\linewidth]{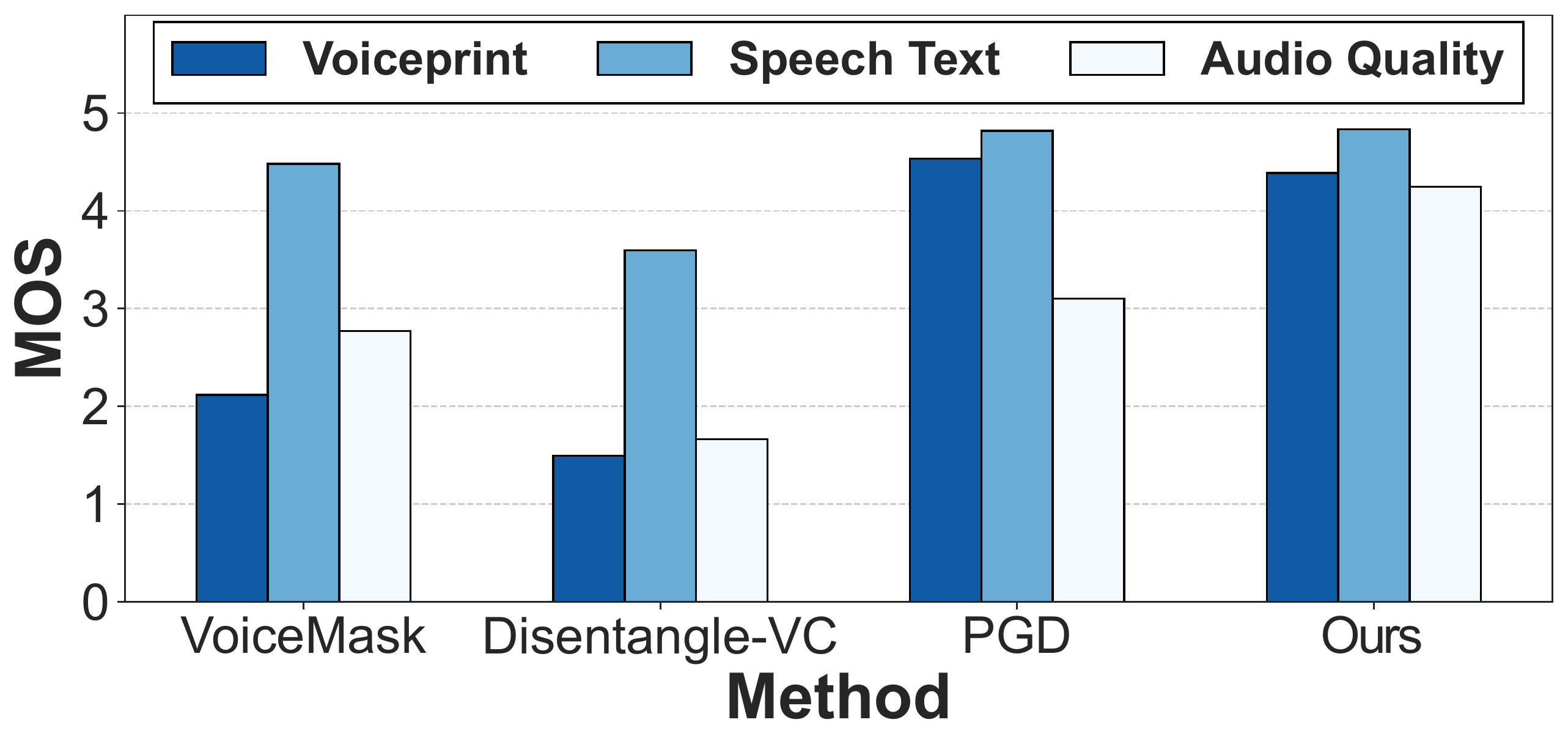}
    \caption{MOS of our system and other three methods.}
    \label{fig:evaluation-mos}
\end{figure}

\subsection{Overall Performance}
We first evaluate the overall effectiveness of our system in terms of voice de-identification and speech preservation, as well as audio distortion. In the experiment, we randomly sample speaker embeddings conditioned on different targets through the pre-trained $\beta$-CVAE decoder and construct 2,189 adversarial examples of the 40 users for each ASI.
These adversarial examples are fed to the four ASIs and an end-to-end ASR pre-trained by SpeechBrain for speaker identification and speech recognition respectively. Considering the convolution operation significantly changes the waveform, we calculate the MCD between the original voices with raw RIRs and de-identified voices with adversarial RIRs.
Table \ref{tab:overall} shows DSRs, WAs, and MCDs of the adversarial examples on the four ASIs. We can observe that the ASIs achieve excellent identification performance on the original voices, but are deceived by our system with high DSRs among 98\%$\sim$100\%, which demonstrates the effectiveness of our de-identification with high confidence. Meanwhile, compared with the original WA of 97.23\% on raw voices, only a 1\%$\sim$4\% drop is introduced into the speech recognition after de-identification, indicating the subtle impact of our adversarial perturbations on the linguistic content utility. Moreover, compared to the typical additive adversarial example methods, our system realizes a lower MCD of 3.7dB$\sim$5.5dB, validating the superior perceptual quality of our RIR-like convolutional adversarial perturbations.

In addition, we further compare our system to SOTA work in voice de-identification, including the frequency warping-based VoiceMask \cite{qian2018h}, disentangled representation-based VC \cite{chouL2019} as well as a typical additive adversarial example method PGD \cite{madry2018t}. We apply these methods to generate de-identified voices for the 40 users, where the warping factor $|\alpha|$ is sampled from [0.08,0.10] as recommended in VoiceMask, random targets are selected for VC, and the perturbation budget size is 0.002 for PGD. As a result, we find all these methods achieve high DSR over 95\% on X-Vector.
Moreover, we also compare their performance in terms of speech preservation and audio distortion. As shown in Figure~\ref{fig:sota-comparison}, VoiceMask and disentangle-VC lead to 8.7\% and 14.3\% WA drop as well as 7.2dB and 11.9dB MCD due to the intrusive voice synthesis, mitigating the utility of voice services. PGD has the least impact on WA but yields an unacceptable MCD over 15dB, indicating the severe distortion of additive perturbations.
Instead, our system achieves a slight WA drop of 4.13\% and a low MCD of 5.45dB with RIR-like convolutional perturbations, thus realizing better speech integrity and less audio distortion.

\begin{figure*}[t]
	\centering
	\subfigure[Original voice.]{
		\includegraphics[width=0.31\linewidth]{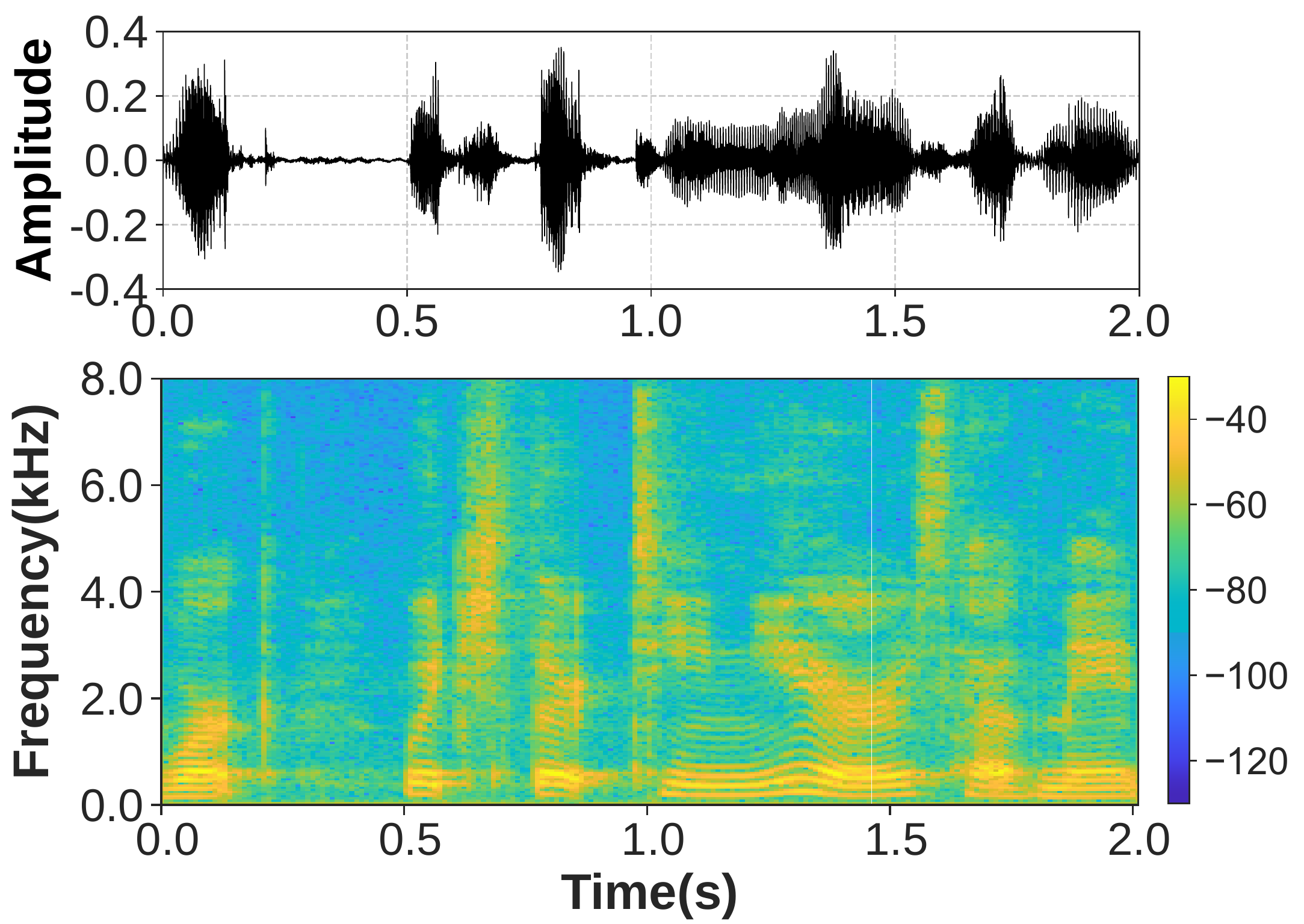}
		\label{fig:original-voice-spec}
	}
	\subfigure[Original voice with raw RIR.]{
		\includegraphics[width=0.31\linewidth]{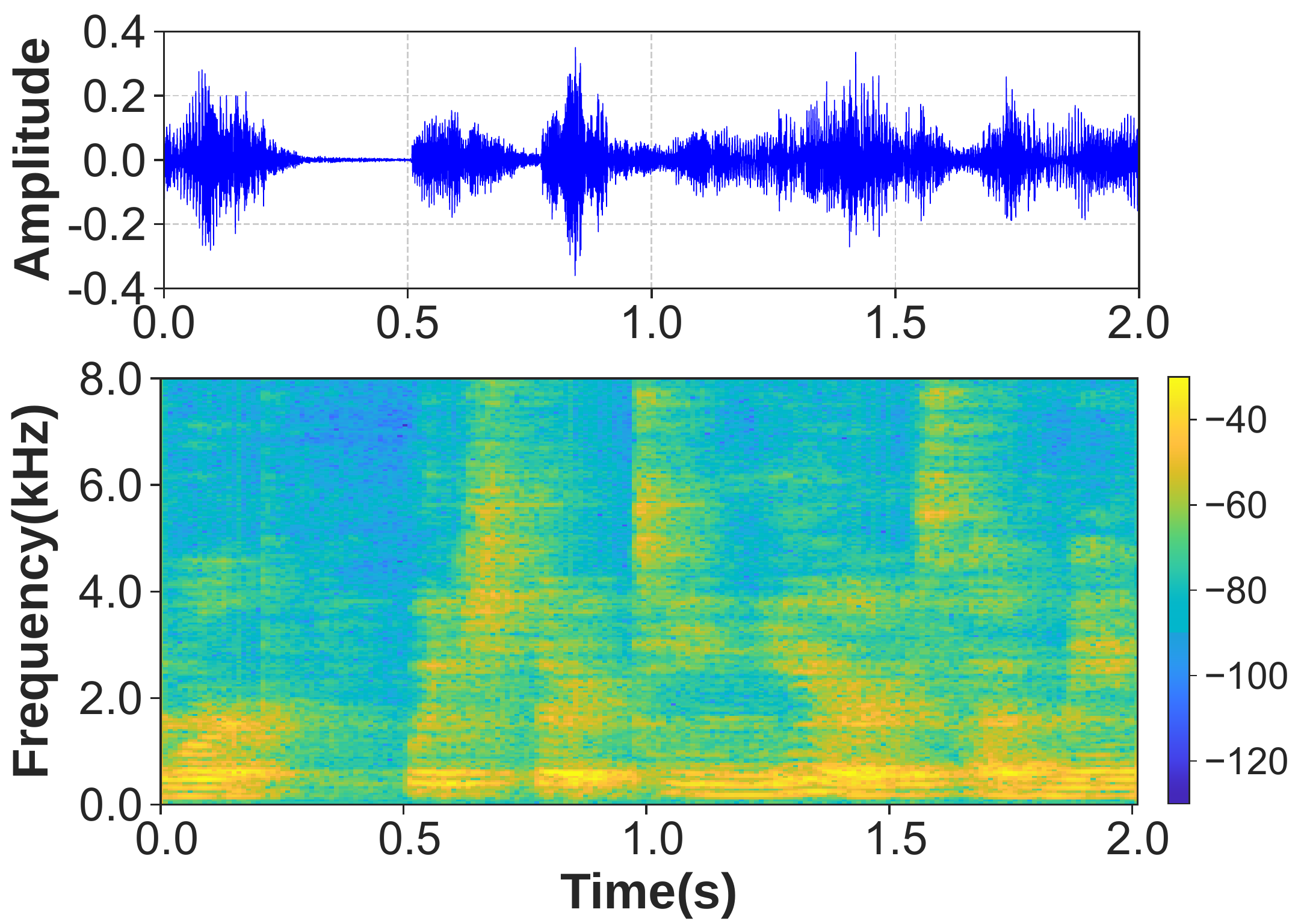}
		\label{fig:benign-example-spec}
	}
	\subfigure[De-identified voice with adversarial RIR.]{
		\includegraphics[width=0.31\linewidth]{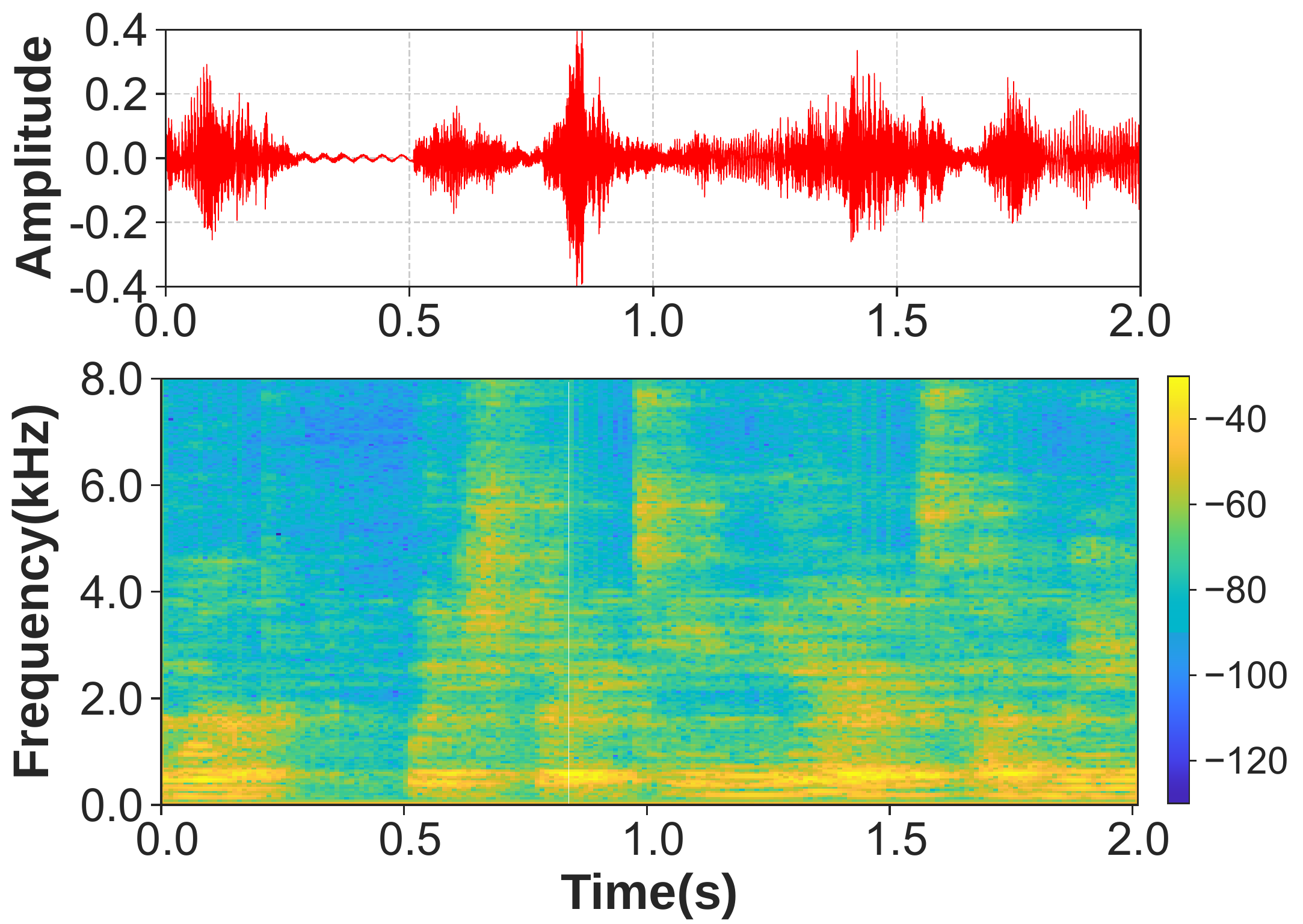}
		\label{fig:adversarial-example-spec}
	}
	\caption{Example of original voice, original voice with raw RIR, and de-identified voice with adversarial RIR.}
	\label{fig:spec-example}
\end{figure*}

\subsection{Performance on Non-Intrusiveness}
Since the subjective metrics (e.g., WA, MCD) can not fully reflect the real perception of human beings,
we further conduct subjective experiments to evaluate the non-intrusiveness of our system to human perception. We recruit 50 volunteers (28 males and 22 females) aged 18$\sim$53, who have no hearing disease and are unaware of the specific de-identification techniques. Note that all the subjective experiments on volunteers are validated by the Institutional Review Board (IRB) in our university. We have these volunteers participate in our MOS test, which includes a comparing trial and a distinguishing trial.

In the comparing trial, the volunteers are asked to listen to 10 pairs of original voices and the corresponding de-identified voices from VoiceMask, Disentangle-VC, PGD, and our system respectively, and then report their intuitive sense of the voice similarity in terms of text, voiceprint, and quality respectively. The volunteers' opinions are recorded as 5-level MOS according to the corresponding descriptions, as shown in Table~\ref{tab:mos-com} (Appendix~\ref{appendix-mos}).
As shown in Figure~\ref{fig:evaluation-mos}, we can see that all these four methods exhibit satisfactory MOS in terms of speech text, but VoiceMask and Disentangle-VC result in poor MOS of 2.12 and 1.50 in terms of voiceprint, due to the significant voiceprint change of voice conversion schemes. Instead, PGD and our system yield high MOS of 4.53 and 4.39 in terms of voiceprint, validating the superiority of adversarial example for remaining voiceprint consistency. Additionally, our system outperforms PGD 0.64 MOS in terms of audio quality, further demonstrating the excellent perceptual quality of our convolutional adversarial perturbations.

\begin{table}[b]
	\centering
	\caption{Results of user study in the distinguishing trial.}
	\label{tab:evaluation-dis}
	\setlength\tabcolsep{2pt}
	\begin{tabular}{ccccc}
		\toprule
		\textbf{Metric} & \textbf{VoiceMask} & \textbf{Disentagle-VC} & \textbf{PGD} & \textbf{Ours} \\
		\midrule
		Is Original(\%) & 4.87 & 0.00 & 34.14 & 45.23 \\
		Unnatural Voiceprint(\%) & \textbf{85.36} & \textbf{79.27} & 4.87 & 11.26 \\
		Illegible Text(\%) & 7.31 & \textbf{60.97} & 0.00 & 3.65 \\
		Distorted Quality(\%) & 31.70 & \textbf{53.66} & \textbf{51.80} & 18.57 \\
		Obvious Reverb(\%) & 19.51 & 24.39 & 15.85 & 39.43 \\
		Other Reason(\%) & 0.00 & 0.81 & \textbf{32.73} & 0.00 \\
		\bottomrule
	\end{tabular}
\end{table}

In the distinguishing trial, we first play an original voice for volunteers to refresh their impression and then play 10 original voices and 10 de-identified voices from VoiceMask, Disentangle-VC, PGD, and our system in random order. For each voice, the volunteers need to determine whether it is original or not. If they regard the voice as not original, they are further asked to choose a reason for this from several options: \textit{unnatural voiceprint}, \textit{illegible text}, \textit{distorted quality}, \textit{obvious reverb} or provide any \textit{other reason} supporting their judgement, as illustrated in Table~\ref{tab:mos-dis} (Appendix~\ref{appendix-mos}).
Table~\ref{tab:evaluation-dis} summarizes the results. Over 45\% de-identified voices from our system are regarded as original. Considering that even 13\% of the original voices are recognized incorrectly, this result is satisfactory and much better than the other three methods. Among the voices that are not recognized as original, 85.36\% of the voices from VoiceMask and 79.27\% from Disentangle-VC are considered unnatural voiceprint, degrading the voiceprint consistency for user experience. PGD performs well in voiceprint and text preservation but over 50\% of voices are considered distorted in quality, and volunteers also reported that they could hear a distinct electrical noise in 32.73\% of voices from PGD in the \textit{other reason} option. Instead, our system well balances the voiceprint, text, and quality with only a natural reverberation effect that does not hurt the user experience.

To explicitly illustrate the non-intrusive de-identification of our system, we present a visual comparison between original voices and adversarial examples in Figure~\ref{fig:spec-example}.
Compared with the original voice in Figure~\ref{fig:original-voice-spec}, Figure~\ref{fig:benign-example-spec} exhibits a distinct waveform and visible drifts in the spectrum, due to the multi-math effect after convolving with the raw RIR. This presents the natural distortion of voice during over-the-air propagation, which is perceived as sound reverberation and does not affect the perceptual quality very much. Meanwhile, we can observe similar results in the de-identified voice in Figure~\ref{fig:adversarial-example-spec} without high-frequency or high-energy irregular artifacts. Hence, the RIR-like convolutional perturbations could maintain natural audibility and provide a better experience for human participants.

\begin{table*}[t]
	\centering
	\caption{Effectiveness under ignorant, semi-informed and informed attacks.}
	\label{tab:adaptiveness}
	\renewcommand{\arraystretch}{1.3}
	\begin{threeparttable}
		\begin{tabular}{cccc>{\columncolor{gray!20}}cc>{\columncolor{gray!20}}cc>{\columncolor{gray!20}}cc>{\columncolor{gray!20}}c}
			\toprule
			\textbf{Setting} & \textbf{Method} & \textbf{Parameter} & \multicolumn{2}{c}{\textbf{D-Vector}} & \multicolumn{2}{c}{\textbf{DeepSpeaker}} & \multicolumn{2}{c}{\textbf{X-Vector}} & \multicolumn{2}{c}{\textbf{Ecapa-TDNN}} \\
			\midrule
			\textbf{Ignorant} & None & None & 8.05 & 99.68 & 7.11 & 100.00 & 5.42 & 98.51 & 1.25 & 99.29 \\
			\hline
			\multirow{4}{*}{\textbf{Semi-informed}} & Bandpass Filter & 200Hz$\sim$7kHz & 8.98 & 70.24 & 8.26 & 78.47 & 5.52 & 97.52 & 1.82 & 99.01\\
			& Re-quantization & 8 bits & 8.20 & 74.59 & 8.67 & 89.34 & 6.37 & 95.19 & 1.89 & 96.35\\
			& Mel Reversion & 80 bins & 11.68 & 60.42 & 8.93 & 74.86 & 5.72 & 98.08 & 2.60 & 88.66\\
			& Psychoacoustic Filter & $\Phi=0$dB & 16.26 & 97.83 & 14.07 & 99.19 & 5.90 & 98.23 & 1.81 & 98.81\\
			\hline
			\multirow{3}{*}{\textbf{Informed}} & Speaker Re-identification & Same as our system & 90.35 & 93.38 & 95.40 & 97.88 & 96.78 & 98.07 & 89.52 & 93.57 \\
			\cline{4-11}
			& TD-based detection (Speech) & Same as our system & \multicolumn{2}{c}{AUC=0.5811} & \multicolumn{2}{c}{AUC=0.5739} & \multicolumn{2}{c}{AUC=0.6158} & \multicolumn{2}{c}{AUC=0.6888} \\
			& TD-based detection (Speaker) & Same as our system & \multicolumn{2}{c}{AUC=0.6064} & \multicolumn{2}{c}{AUC=0.6331} & \multicolumn{2}{c}{AUC=0.5923} & \multicolumn{2}{c}{AUC=0.5313} \\
			\bottomrule
		\end{tabular}
		\begin{tablenotes}
			\footnotesize
			\item[1] DSRs on the original voices and de-identified voices are marked with white and gray backgrounds respectively.
			\item[2] AUC refers to the perturbation detection performance on the de-identified voices under the best split ratio.
		\end{tablenotes}
	\end{threeparttable}
	\renewcommand{\arraystretch}{1}
\end{table*}

\begin{figure*}[t]
	\centering
	\subfigure[Original.]{
		\includegraphics[width=0.235\linewidth]{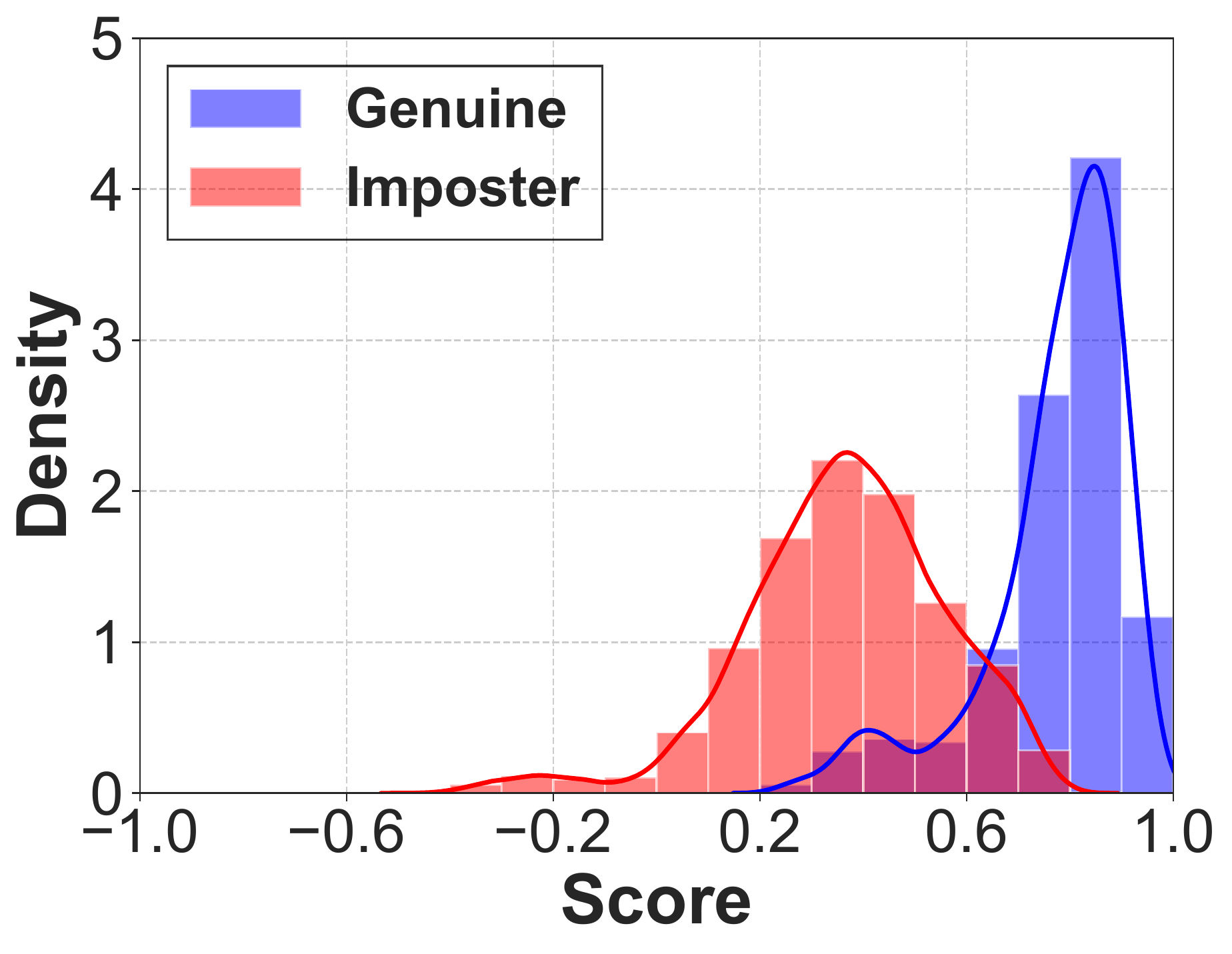}
		\label{fig:beni-xvect}
	}
	\hspace{-3mm}
	\subfigure[Ignorant.]{
		\includegraphics[width=0.235\linewidth]{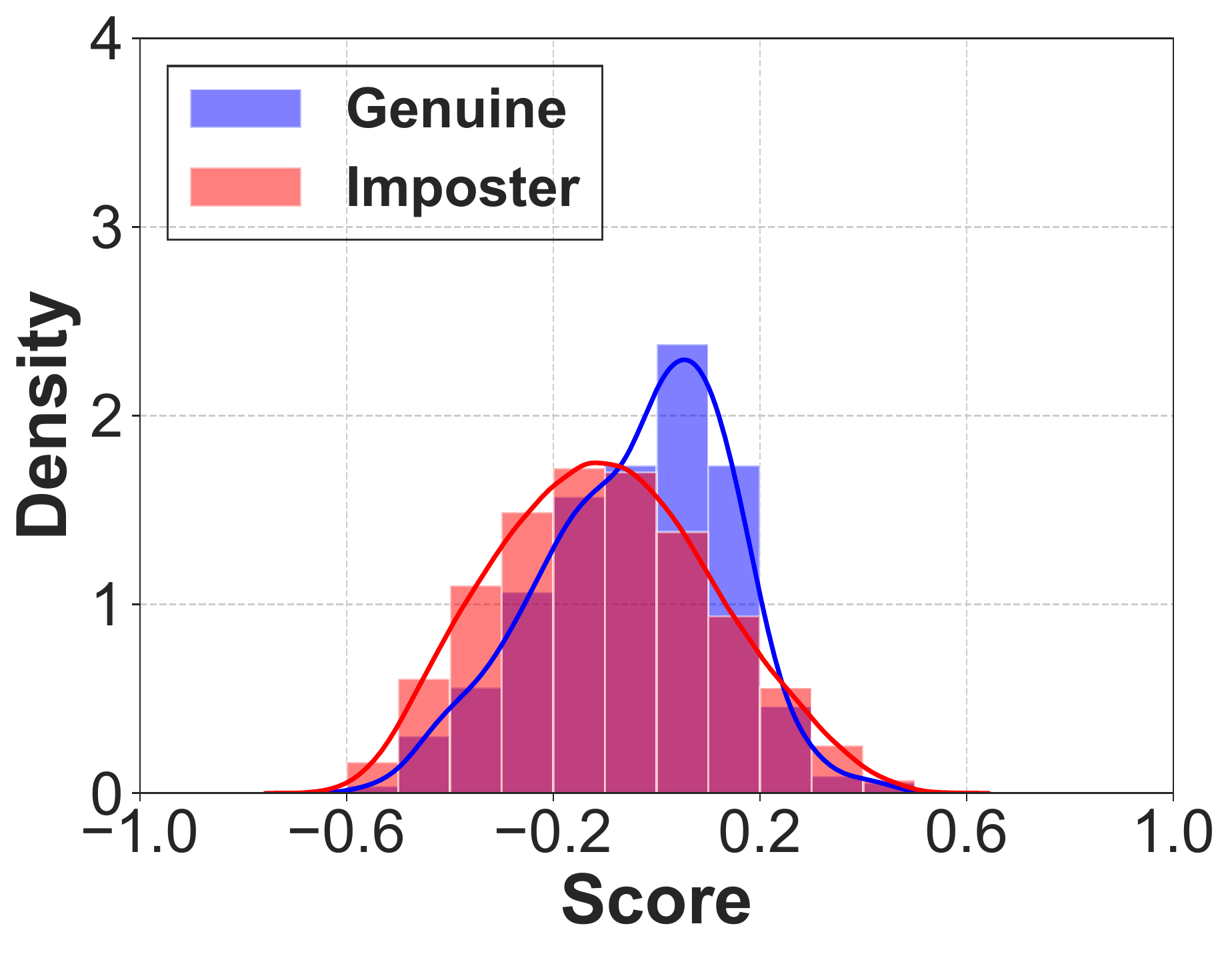}
		\label{fig:igno-xvect}
	}
	\hspace{-3mm}
	\subfigure[Semi-informed.]{
		\includegraphics[width=0.235\linewidth]{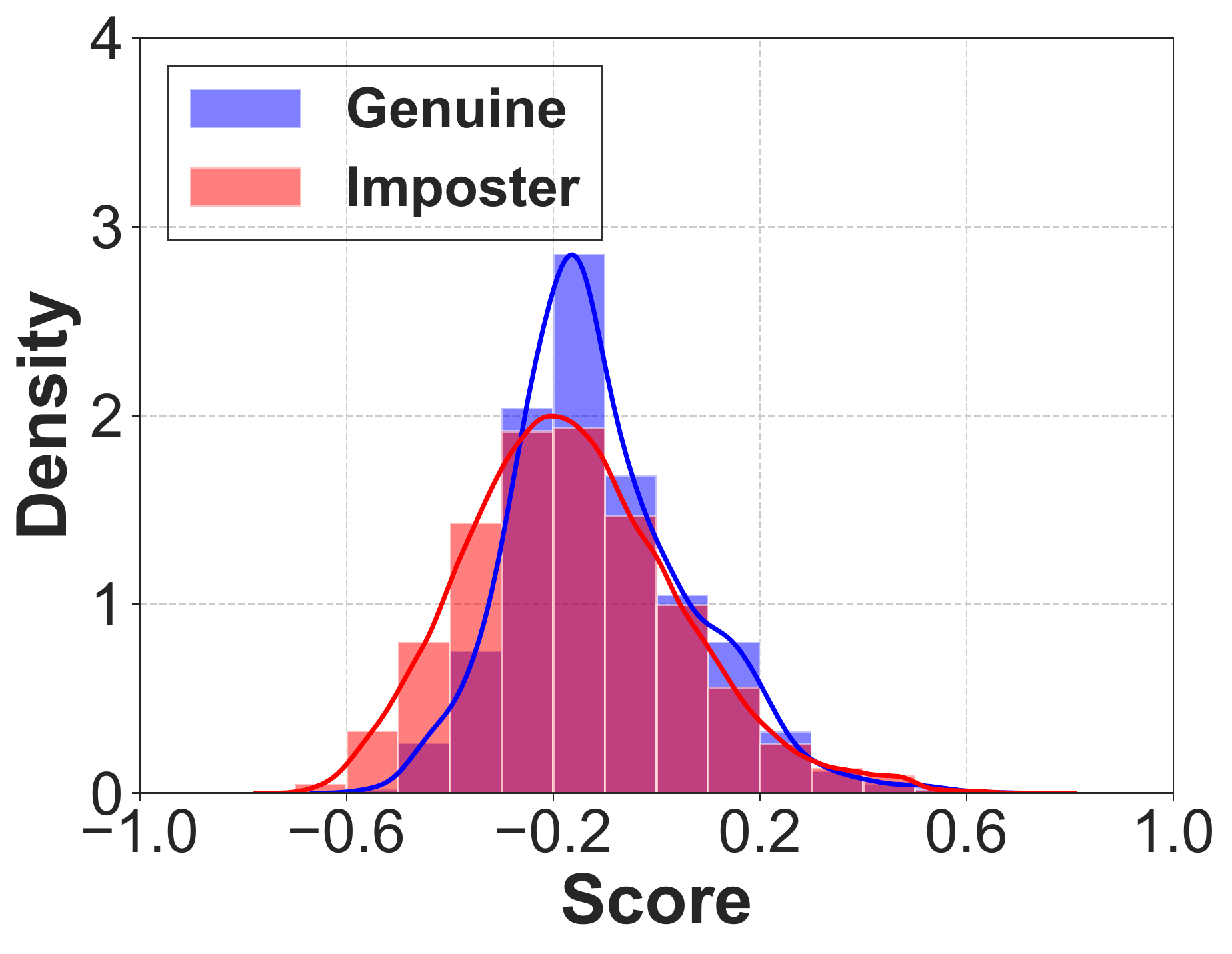}
		\label{fig:semi-xvect}
	}
	\hspace{-3mm}
	\subfigure[Informed.]{
		\includegraphics[width=0.235\linewidth]{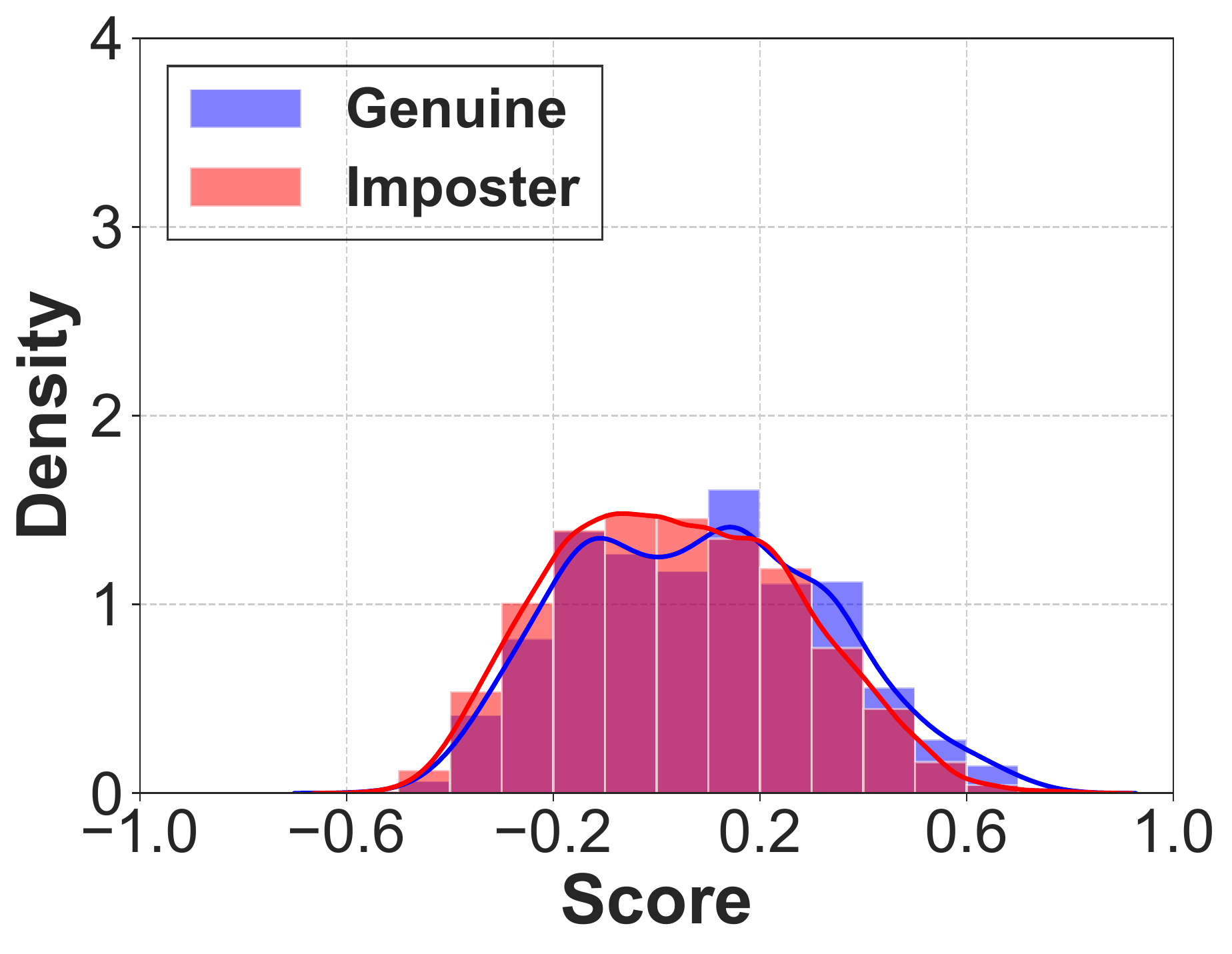}
		\label{fig:info-xvect}
	}
	\caption{Score distribution of genuine and imposter trials on original and de-identified voices under the ignorant, semi-informed and informed settings.}
	\label{fig:score-distribution}
\end{figure*}

\subsection{Performance on Adaptiveness}
As a privacy protection solution for voice service users, our system is required to be adaptive for resisting not only powerful ASIs but also skilled service providers. These providers could introduce specialized resistance methods based on the prior knowledge of our de-identification strategy, exhibiting higher difficulty for identity preserving.
Hence, we further evaluate the adaptiveness of our system under three different kinds of attacks.
The three kinds of attacks are categorized based on their prior knowledge of our de-identification strategy, i.e.,
(1) \textit{Ignorant Attack}: the adversary is unaware of the de-identification, and directly performs ASI on the de-identified voices. (2) \textit{Semi-informed Attack}: the adversary understands the adversarial example-based de-identification strategy but without implementation details, in which the adversary tries to corrupt the adversarial perturbations before applying ASI. (3) \textit{Informed Attack}: the adversary has full knowledge about the de-identification strategy and model details, where the substitute model and data could be obtained by the adversary for speaker re-identification and perturbation detection.

Since the ignorant adversary applies no countermeasure to resist our system, we focus on implementing the semi-informed and informed attacks in this experiment.
Specifically, we implement four common signal processing approaches for perturbation disruption as the semi-informed attacks, i.e., (1) \textit{Bandpass Filtering} discards the frequency bands higher than 7kHz and lower than 200Hz; (2) \textit{Re-quantization} quantizes the voice from 16 bits to 8 bits and then de-quantizes it back to 16 bits; (3) \textit{Mel Re-transform} converts the voice from waveform to 80-bin MFCCs and then transforms it back to waveform inversely; (4) \textit{Psychoacoustic Filter} eliminates the frequency components 0dB higher than the hearing threshold. All the parameters are configured to produce maximum distortion while remaining the normal speaker identification performance according to WaveGuard \cite{hussainNDMK21} and Dompteur \cite{Eisenhofer2021}. As for informed attacks, we implement three advanced attacks: (1) \textit{Speaker Re-identification}: the service provider performs the same de-identification on each user's enrollment voices for re-identification as introduced in \cite{mohan2020e}; 
(2) \textit{Temporal Dependency-based (TD-based) Perturbation Detection}: the service provider discriminates adversarial perturbations through exploring the speech text or speaker identity change between the partial and full utterances \cite{yangLCS19}, which is specifically designed to detect audio adversarial examples.

\begin{figure}[t]
	\centering
	\subfigure[Speech Consistency.]{
		\includegraphics[width=0.45\linewidth]{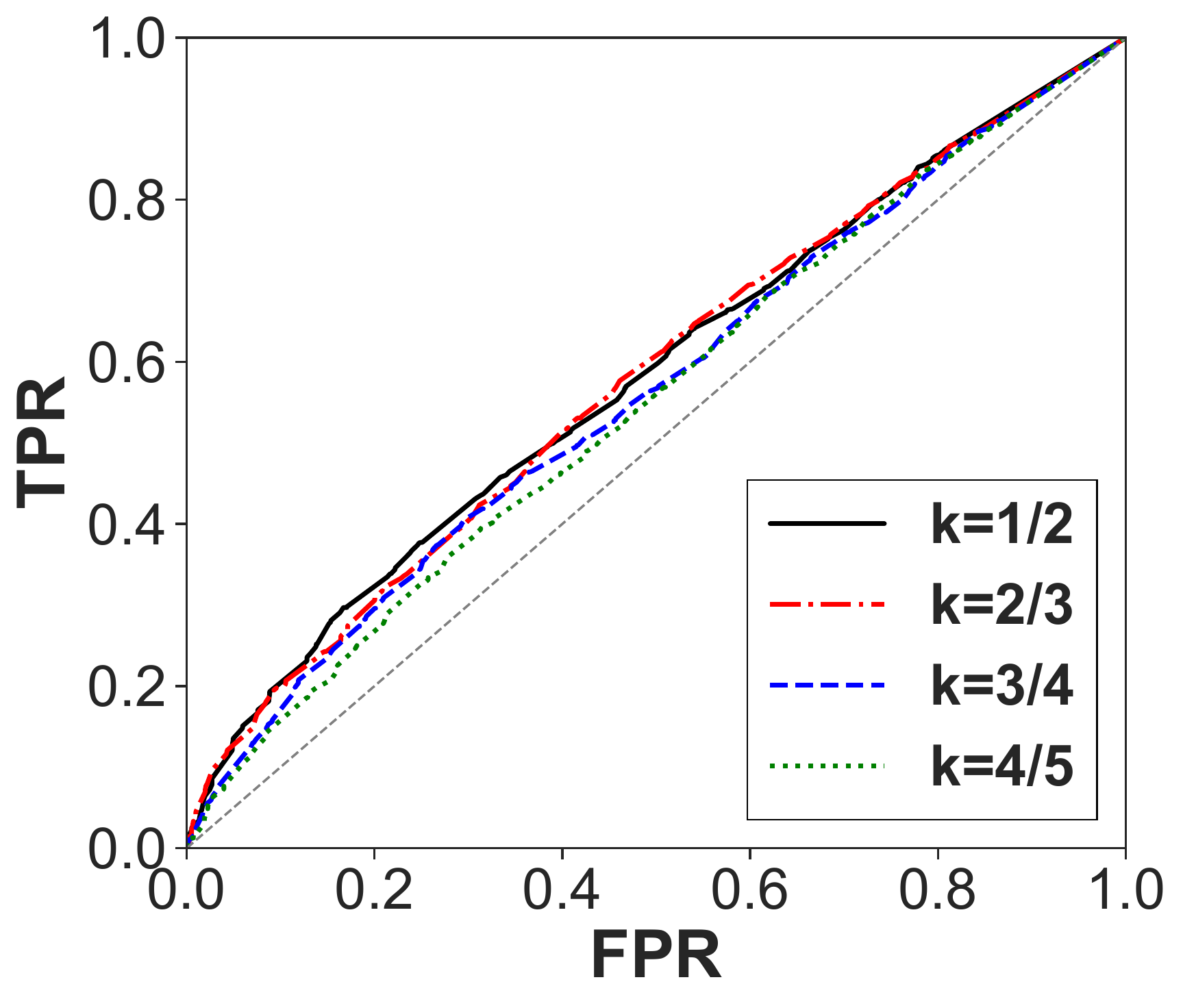}
		\label{fig:td-roc-speech}
	}
	\subfigure[Speaker Consistency.]{
		\includegraphics[width=0.45\linewidth]{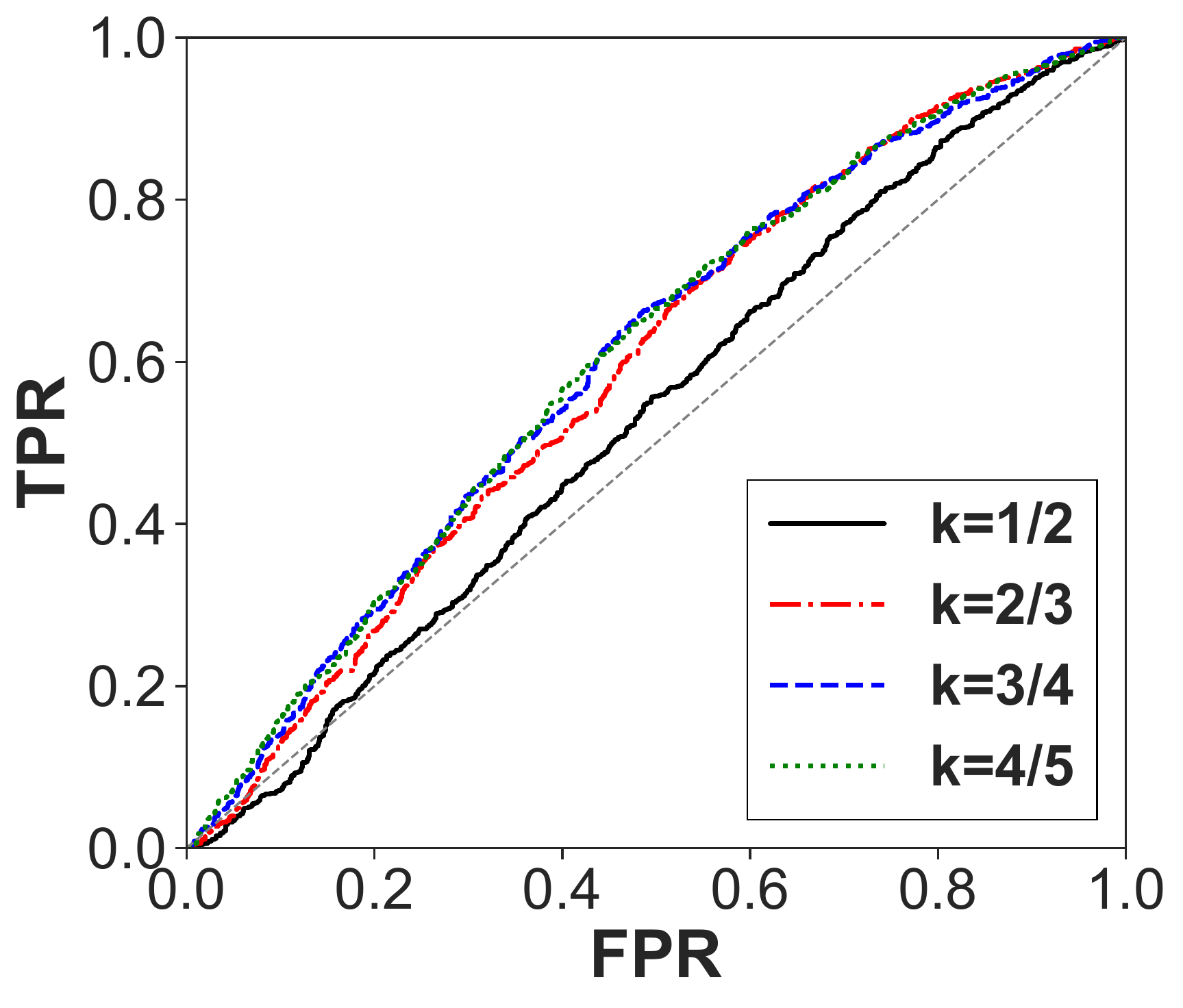}
		\label{fig:td-roc-speaker}
	}
	\caption{ROC curves of TD-based perturbation detection over split ratios.}
	\label{fig:td-roc}
\end{figure}

Table~\ref{tab:adaptiveness} summarizes the results. In the ignorant setting, all the four ASIs with excellent identification performance are deceived by more than 98\%. In the semi-informed setting, all these signal processing approaches have limited impact on the de-identification performance of our system, especially on the de-identified voices generated against powerful ASIs (e.g., X-Vector and Ecapa-TDNN), indicating the robustness of our convolutional adversarial perturbations. In the informed setting, both the DSR on original and de-identified voices reaches over 90\% after speaker re-identification. This is because the de-identified enrollment voices have diverse identities that are outside the original enrollment set. In the TD-based detection, we compare the WA and DSR of partial and full de-identified voices over different split ratios to determine the threshold for perturbation detection. As shown in Figure~\ref{fig:td-roc}, the ROC curves of TD-based detection are close to the curve of the random classifier, and the AUCs with the best split ratio on these four ASIs are 0.53$\sim$0.69, showing their poor perturbation detection performance and the robustness of our perturbations with respect to speech and speaker change.

\begin{table}[b]
	\centering
	\caption{DSR of ensembled model across different ASIs.}
	\label{tab:ensemble}
	\setlength\tabcolsep{3pt}
	\begin{tabular}{ccc|cc}
		\toprule
		\multicolumn{3}{c|}{\textbf{Substitute ASIs}} & \textbf{Target ASI} & \textbf{DSR(\%)} \\
		\midrule
		DeepSpeaker & X-Vector & Ecapa-TDNN & D-Vector & 88.77$\pm$3.38\\
		D-Vector & X-Vector & Ecapa-TDNN & DeepSpeaker & 95.80$\pm$2.42\\
		D-Vector & DeepSpeaker & Ecapa-TDNN & X-Vector & 92.81$\pm$4.00\\
		D-Vector & DeepSpeaker & X-Vector & Ecapa-TDNN & 69.18$\pm$9.86\\
		\bottomrule
	\end{tabular}
\end{table}

\begin{figure}[t]
	\centering
	\subfigure[Conditional Sampling.]{
		\includegraphics[height=2.9cm]{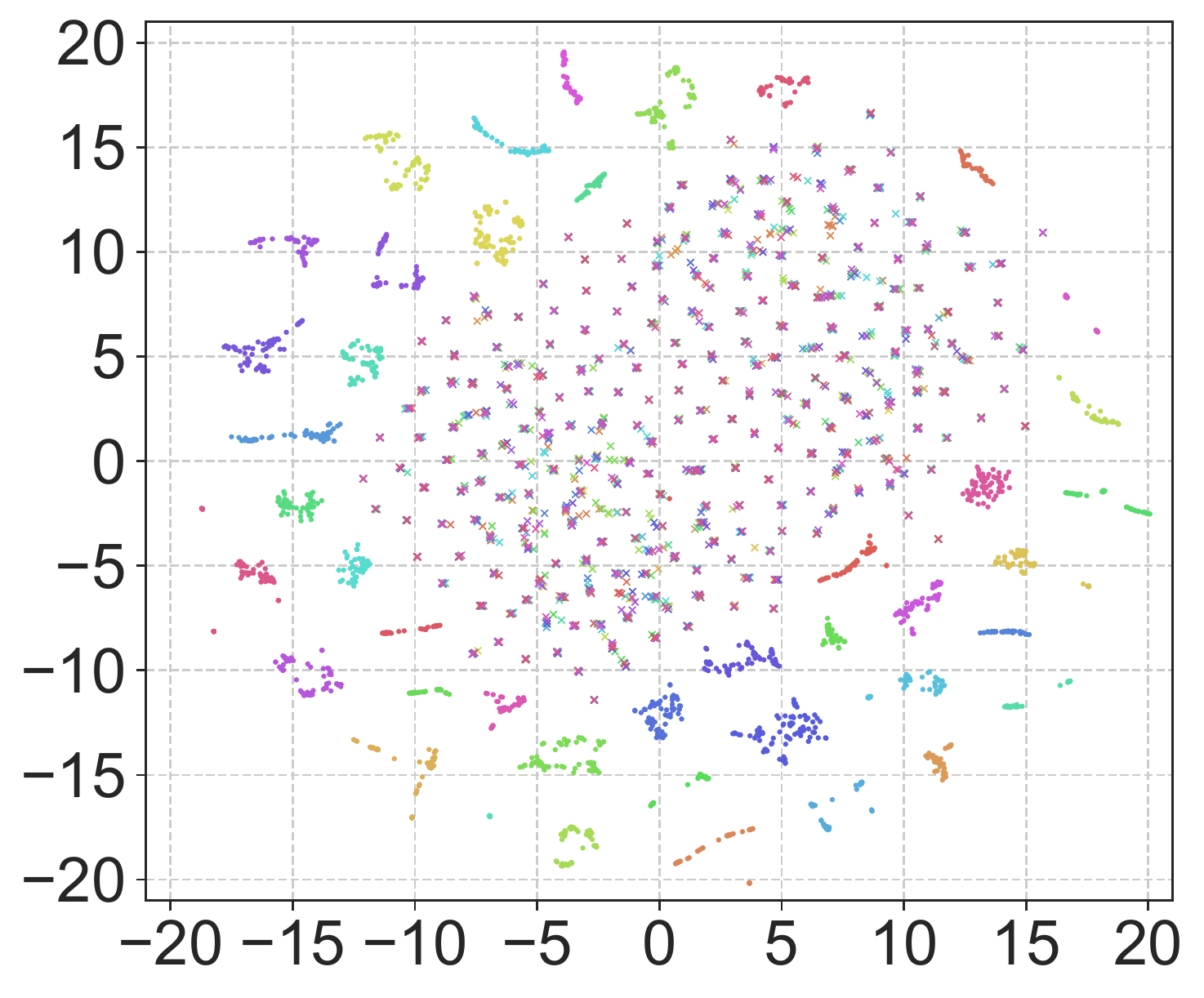}
		\label{fig:emb-tsne-sample}
	}
	\hspace{-2mm}
	\subfigure[Semantic Interpolation.]{
		\includegraphics[height=2.9cm]{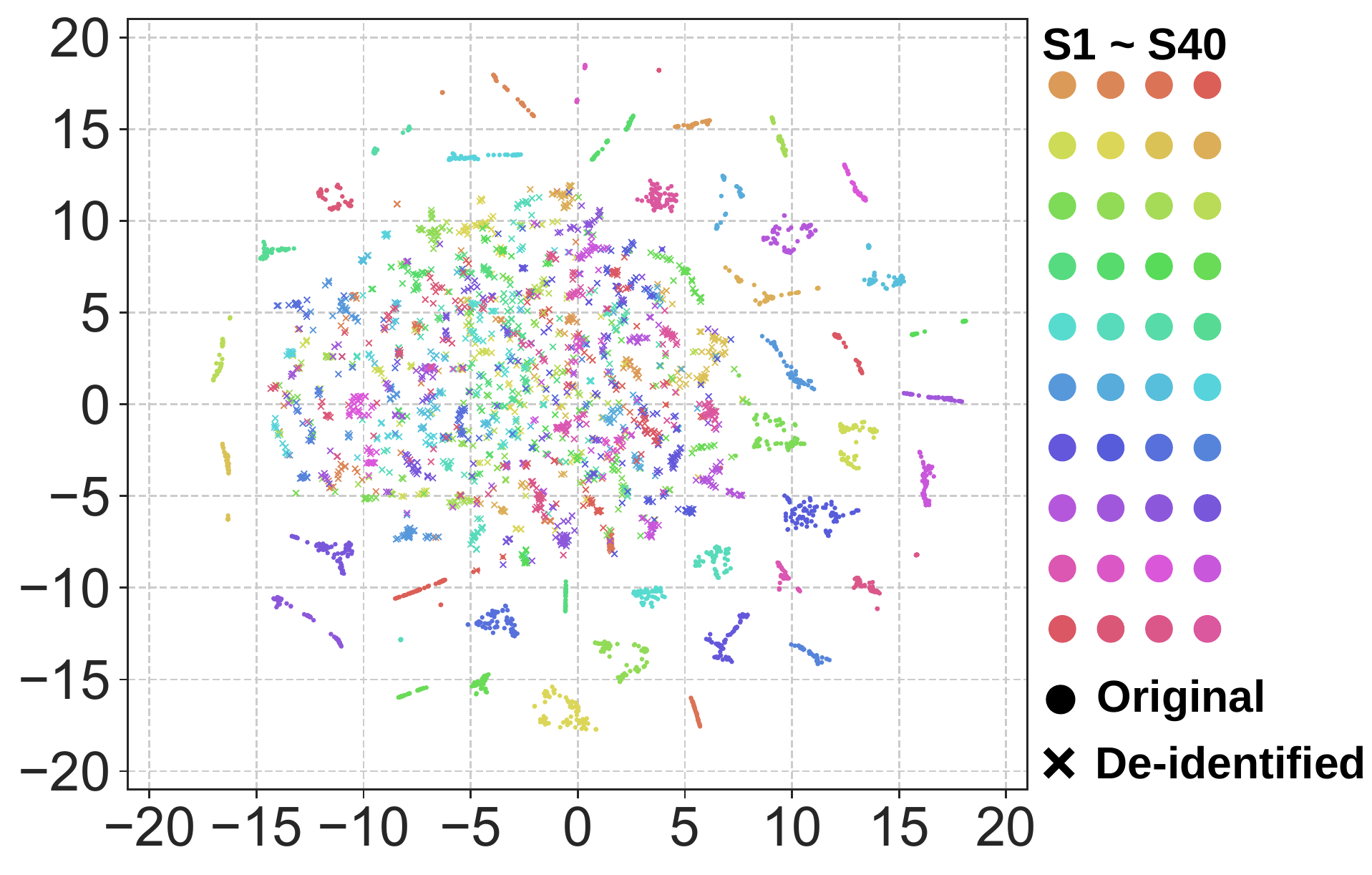}
		\label{fig:emb-tsne-inter}
	}
	\caption{T-SNE visualization of speaker embeddings of original and de-identified voices through target sampling and interpolation.}
	\label{fig:tsne-example}
\end{figure}

To better understand the adaptiveness of our system, we present the score distribution of \textit{Genuine} and \textit{Imposter} trials on the original and de-identified voices, i.e., the log-likelihood ratios between same-speaker and different-speaker hypotheses \cite{mohan2020e}. As shown in Figure~\ref{fig:score-distribution}, the genuine and imposter score distribution of the original voices are separated from each other, so that the ASI can distinguish them. But after our de-identification, the score distribution of genuine and imposter trials overlap under all the three settings, thus hard to be distinguished and linked by advanced service providers.
In addition, we also apply t-SNE to visualize the speaker embeddings of the original and de-identified voices from the 40 users to investigate the diversity. As shown in Figure~\ref{fig:tsne-example}, the de-identified voices are far away from the original voices and mapped to diverse targets instead of clustering together. Moreover, compared to conditional sampling, the de-identified voices exhibit more diversity in semantic interpolation. Hence, our system can realize diverse identity transformation and successful de-identification even under the informed setting.

\subsection{Performance on Transferability}
In practice, the adversary usually has no prior knowledge of the implementation details of target ASIs. Fortunately, the adversarial example has been demonstrated to be able to generalize on other models with similar tasks. Hence, we also evaluate the transferability of our system on unseen models. In each experiment, we select one of the four ASI models as the substitute system to construct adversarial examples, which are then fed to the other three ASI models as targets respectively, for performance evaluation.
By analogy, the experiment is repeated on the four ASI models in turn.

\begin{table}[b]
	\centering
	\caption{DSR on commercial systems.}
	\label{tab:commercial}
	\setlength\tabcolsep{2pt}
	\begin{tabular}{cccc}
		\toprule
		\textbf{System} & \textbf{Threshold} & \textbf{Original DSR(\%)} & \textbf{De-identified DSR(\%)} \\
		\midrule
		iFLYTEK & 0.65 &  3.55$\pm$1.79 & 83.22$\pm$6.06 \\
		Microsoft Azure & 0.45 & 1.88$\pm$0.82 & 76.74$\pm$11.25 \\
		\bottomrule
	\end{tabular}
\end{table}

\begin{figure}[t]
	\centering
	\begin{minipage}{0.49\linewidth}
		\centering
        \includegraphics[width=\linewidth]{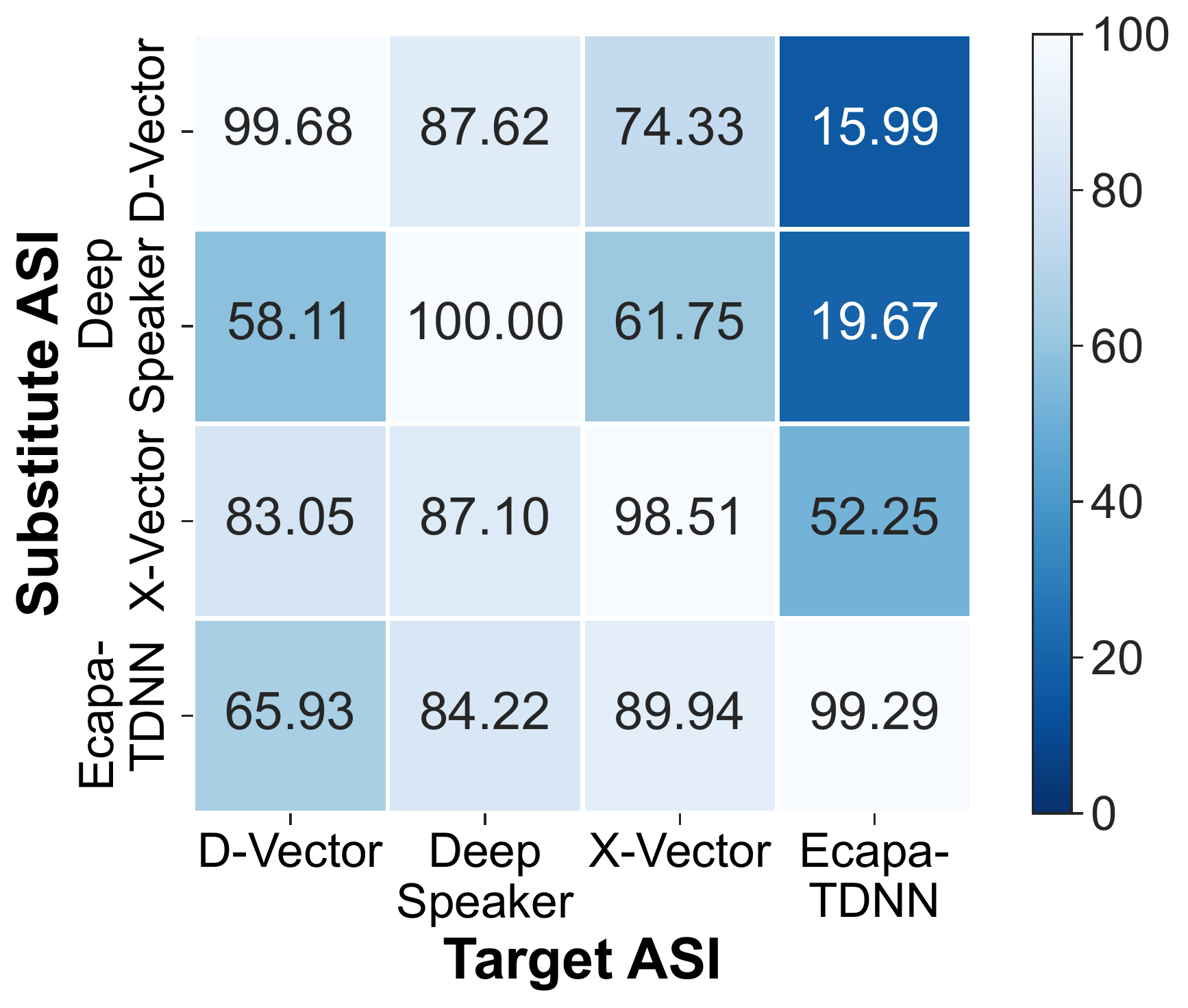}
        \vspace{-2mm}
    	\caption{DSR across different ASI systems.}
    	\label{fig:evaluation-confusion-matrix}
	\end{minipage}
	\begin{minipage}{0.49\linewidth}
	    \centering
    	\includegraphics[width=\linewidth]{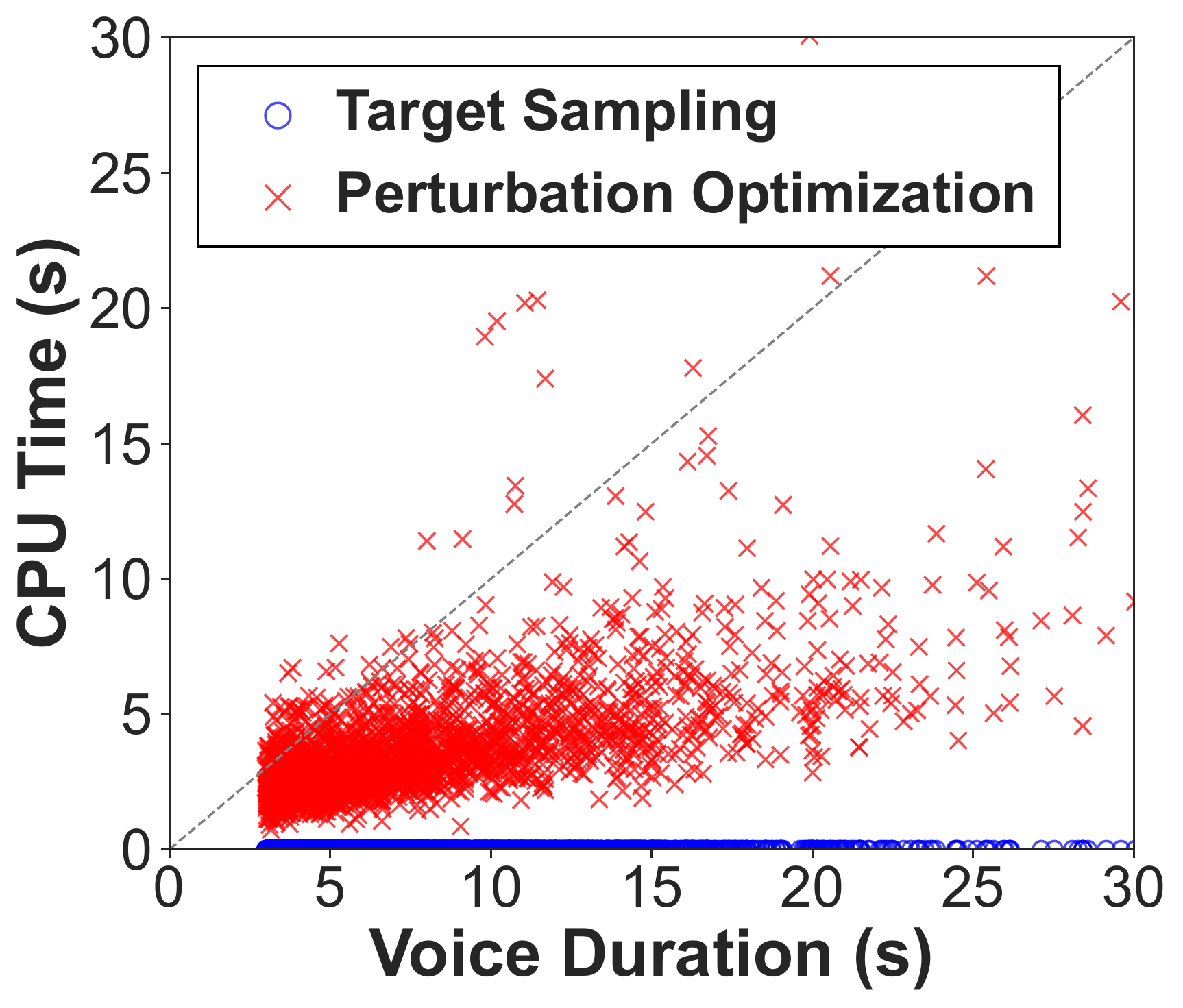}
    	\vspace{-2mm}
    	\caption{Time cost of target sampling and perturbation optimization.}
    	\label{fig:time-cost}
	\end{minipage}
	\vspace{-3mm}
\end{figure}

Figure~\ref{fig:evaluation-confusion-matrix} shows the DSR confusion matrix of our system across different ASIs.
We can observe that our system exhibits variable transferability between different substitute and target ASIs, with DSRs varying from 15.99\% to 89.94\%. This is because the single model for training can only provide limited knowledge for transfer de-identification.
We further perform the ensemble learning to enhance the generalization ability of adversarial examples, which integrates several ASIs as substitutes for training but fetches a single one out as the target system for evaluation. As shown in Table~\ref{tab:ensemble}, with the ensemble learning, its DSRs increase to 88.77\%, 95.80\%, 92.81\% and 69.18\% on four unseen ASIs respectively, indicating the satisfactory transferability of our system.

We also evaluate the transferability of ensembled model on commercial ASI systems, including iFLYTEK \cite{iflyteck} and Microsoft Azure \cite{azure}. We query these commercial systems for enrollment and evaluation via the given HTTP REST API and receive the identification feedback. In each experiment, the same 40 users are enrolled to create speaker profiles, and then the rest 1,289 utterances are used to evaluate their normal identification performance and calculate the thresholds. After that, they are further used to generate adversarial examples on the substitute system ensembling the four aforementioned ASIs. As shown in Table \ref{tab:commercial}, our system could deceive iFLYTEK and Microsoft Azure with average DSRs of 83.22\% and 76.74\%, respectively. Compared with the DSR of 3.55\% and 1.88\% on the original voices, such a result indicates a strong identity protection capability of our system. But it can be also observed that the standard deviations of their DSRs are 6\%$\sim$11\%. This is because the model details of commercial systems are completely agnostic to us, resulting in the difficulty to generalize on them with several dissimilar substitute ASIs. Compared to the SOTA work VoiceMask \cite{qian2018h}, our system scarifies partial de-identification performance (3.26\%) on commercial systems but achieves much better voice utility, thus providing a balanced privacy-utility trade-off for voice services.

\subsection{Computation and Storage Overhead}
Considering the real-world application of voice services and the practical deployment for our system, we further evaluate the computation and storage overhead on different devices.

Specifically, we record the CPU time and derive average RTRs by iterating 1,000 times. As shown in Table~\ref{tab:rtr}, our system achieves an average RTR of 0.63 on Dell OptiPlex 7080, indicating the real-time performance without extra latency in human sense.
But for Lenovo YOGA C940 and ThinkPad X1 Carbon with much lower computing power, RTR rapidly increases to 1.34 and 1.65.
By further investigating the time cost of different stages in our system as shown in Figure~\ref{fig:time-cost}, we find that the target embedding sampling operation takes up little CPU time, while the perturbation optimization is the most expensive one (over 95\%). This observation verifies the efficiency of our target generation design, and encourages us to optimize a more efficient perturbation construction. Moreover, our current implementation is built on Python, which exhibits low computing efficiency. Once the implementation turns to more low-level language (e.g., C or C++), our system could achieve better real-time performance.
As for the storage cost, our system only occupies 29.1M (including the pre-trained $\beta$-CVAE decoder and substitute ASI model), thanks to the lightweight decoder for target embeddings generation. Compared with most typical target voice pool-based solutions occupying 300M storage for 250 target speakers \cite{han2020v}, \textit{VoiceClock} significantly reduces the storage requirement, thus more suitable for deployment on resource-limited devices.

\begin{table}[b]
	\centering
	\caption{RTR on different devices.}
	\label{tab:rtr}
	\resizebox{\linewidth}{!}{
		\begin{tabular}{cccc}
			\toprule
			\textbf{Device} & \textbf{CPU} & \textbf{Memory} & \textbf{RTR} \\
			\midrule
			Dell OptiPlex 7080 & i7-10700@2.9GHz & 32GB & 0.63$\pm$0.07 \\
			Lenovo YOGA C940 & i7-1065G7@1.3GHz & 16GB & 1.34$\pm$0.06 \\
			ThinkPad X1 Carbon & i7-8550U@1.80GHz & 16GB & 1.65$\pm$0.11 \\
			\bottomrule
	\end{tabular}}
\end{table}

	\section{Discussion}
\label{sec:discussion}
This section discusses the practical issues of our system.

\textbf{Deployment mode in practice}. Our system acts as a voice input filter at the user side and can be deployed in two modes. The first mode is to integrate our system into the Operating System (OS) as a basic facility for the full control of all voice input, which allows trusted APPs to access raw voices while feeding untrusted ones the de-identified voices only. This ensures a system-level security guarantee of users' voiceprints but requires additional OS modifications.
In the second mode, our system is installed as a voice input plugin on the users' input editor, which performs de-identification every time the voice input interface is invoked. This requires users to install our system in advance and authorize corresponding permissions, for which an open-source version is helpful to release users' privacy concerns.

\textbf{Compatibility with voiceprint authentication}. Despite voiceprint-irrelevant voice services (e.g., speech recognition), users may also rely on voiceprint-based systems, such as WeChat voiceprint lock \cite{wechat} and Siri personalized activation \cite{siri}. In these cases, these users have to offer their clean voiceprints for accurate authentication, but our system may hinder such normal usages.
Hence, a compatibility mechanism between de-identification and authentication is needed. Considering the fixed wake texts for authentication services usually, a straightforward solution is to shut down our system temporarily when detecting these specific texts, and automatically restore the protection as soon as the authentication is finished.
Another solution may ensure more comprehensive voiceprint protection, where we can transform users' voiceprints to the same target speaker for enrollment and later activation or unlocking. This could enable de-identification and authentication simultaneously.

	\section{Related Work}
\label{sec:related-work}
In this section, we review several key related researches of voice de-identification.

\textbf{Signal processing-based voice de-identification.}
Early studies \cite{jin2009v,abou2015a,magarinos2016p,vaidya2019y} on voice de-identification mainly exploit signal processing techniques (e.g., frequency warping, amplitude scaling, duration warping) to modify the spectral and prosodic features for manipulating the in-depth voiceprint.
However, these methods require parallel source and target voices to train the voice transformation, i.e., two voices with the same texts and timestamps, limiting its utility in practice. To address this issue, the following works \cite{pobar2014o,magarinos2017r} propose to pre-calculate a set of voice transformations between multiple pairs of source and target speakers for online de-identification. But the voices produced by this method are filled with obvious artifacts due to the synthetic speakers. Hence, a new paradigm based on Vocal Tract Length Normalization (VTLN) \cite{eide1996a} is proposed to realize voice transformation without parallel corpus. Specifically, VLTN-based works \cite{qian2018h,zhang2020e,qian2018t} stretch or compress the voice spectrum frame by frame according to a warp function, and synthesize de-identified voices through a waveform vocoder. However, the invertible warping function used in these approaches is probably employed to recover the original voice by informed adversaries \cite{srivastava2020e}. This design indicates the intrinsic vulnerability for voice de-identification.

\textbf{Deep learning-based voice de-identification.}
With the advances of artificial intelligence, numerous studies turn to explore voice de-identification based on learning methods. One branch of them \cite{justin2015s,qian2018t,ahmed2020p} integrates speech-to-text and text-to-speech techniques to transform voices to texts and then re-synthesize de-identified voices, which eliminates identify information but introduces unacceptable overhead. Another branch \cite{abs1905,han2020v,mohammadi2017a,srivastava2020d,turner2022g} proposes X-Vector-based voice conversion schemes, which replaces the source embedding with the target one in the latent space of X-Vector.
By means of a vocoder or a neural source filter, these works could synthesize anonymized voices with satisfactory quality. However, a pool of speaker voices needs to be pre-collected, and complex target selection strategies are mandatory to ensure the de-identification performance. These lead to difficulties of practical deployment on users' resource-limited devices. More recent studies \cite{srivastava2019p,abs2011} propose to learn de-identified speech representation using adversarial training at the user side, but require a redesign of existing service architecture, also hindering their deployments.

Although the aforementioned works make great efforts to preserve speaker identity from disclosure, the direct voiceprint modification or speech re-synthesizing would produce a perceptually inconsistent voiceprint with low diversity. This degrades not only the user experience of human participants but also the resistance to informed attacks. Different from these studies, our work aims to design a non-intrusive and adaptive voice de-identification system, which enables any source user to disguise any target speaker with inaudible convolutional adversarial perturbations.

	\section{Conclusion}
\label{sec:conclusion}

This paper presents a voice de-identification system, which turns adversarial examples as a defense tool against Automatic Speaker Identification (ASI) to balance the privacy and utility of voice services. We first conduct preliminary studies to investigate the feasibility of adversarial example-based voice de-identification and find the perceivable distortion and low diversity problems of additive adversarial perturbations. To address these issues, we propose room impulse responses to embed convolutional adversarial perturbations into the natural reverberation, improving the perceptual consistency for human participants. In addition, we design a lightweight conditional variational auto-encoder to generate diverse targets on demand, and then construct input-specific perturbations through a triplet loss architecture, enabling any-to-any identity conversion for adaptive de-identification. Experimental results show that our system could effectively de-identify users on mainstream open-source ASIs and commercial systems, and achieve excellent utility in both voice quality and speech integrity.

    \bibliographystyle{IEEEtran}
    \bibliography{ref}

\begin{thebibliography}{10}
\providecommand{\url}[1]{#1}
\csname url@samestyle\endcsname
\providecommand{\newblock}{\relax}
\providecommand{\bibinfo}[2]{#2}
\providecommand{\BIBentrySTDinterwordspacing}{\spaceskip=0pt\relax}
\providecommand{\BIBentryALTinterwordstretchfactor}{4}
\providecommand{\BIBentryALTinterwordspacing}{\spaceskip=\fontdimen2\font plus
\BIBentryALTinterwordstretchfactor\fontdimen3\font minus
  \fontdimen4\font\relax}
\providecommand{\BIBforeignlanguage}[2]{{%
\expandafter\ifx\csname l@#1\endcsname\relax
\typeout{** WARNING: IEEEtran.bst: No hyphenation pattern has been}%
\typeout{** loaded for the language `#1'. Using the pattern for}%
\typeout{** the default language instead.}%
\else
\language=\csname l@#1\endcsname
\fi
#2}}
\providecommand{\BIBdecl}{\relax}
\BIBdecl

\bibitem{wechat}
{Wechat Official}, ``{Voiceprint: The New WeChat Password},''
  \url{https://blog.wechat.com/2015/05/21/voiceprint-the-new-wechat-password},
  2015.

\bibitem{hsbc}
R.~Bharadwaj, ``Voice and speech recognition in banking – what’s possible
  today,'' \url{https://www.hsbc.com.hk/ways-to-bank/phone/voice-id/}, 2019.

\bibitem{td}
{TD Bank}, ``Td voiceprint,''
  \url{https://www.tdbank.com/bank/tdvoiceprint.html}, 2022.

\bibitem{google}
{Google Privacy \& Terms}, ``{How Google Voice works},''
  \url{https://policies.google.com/technologies/voice?hl=en-US}, 2022.

\bibitem{microsoft}
Microsoft, ``{How does Microsoft protect my privacy while improving its speech
  recognition technology?}''
  \url{https://support.microsoft.com/en-us/windows/how-does-microsoft-protect-my-privacy-while-improving-its-speech-recognition-technology-f465d7a7-4a4f-40b7-9441-f0e6e97e24ec},
  2022.

\bibitem{apple}
Forbes, ``{Apple Just Gave 1.5 Billion iPad, iPhone Users A Reason To Leave},''
  \url{https://www.forbes.com/sites/gordonkelly/2022/02/12/apple-iphone-ipad-siri-audio-recordings-iphone-privacy/?sh=68fc85bd4193},
  2021.

\bibitem{alexa}
{The New York Times}, ``{Amazon’s Alexa Never Stops Listening to You. Should
  You Worry?}''
  \url{https://www.nytimes.com/wirecutter/blog/amazons-alexa-never-stops-listening-to-you/},
  2019.

\bibitem{iflyteck}
{iFLYTEK Open Platform}, ``{Voiceprint Recognition},''
  \url{https://www.xfyun.cn/service/isv}, 2022.

\bibitem{azure}
{Microsoft Azure Congnitive Service}, ``{Speaker recognition},''
  \url{https://azure.microsoft.com/en-us/services/cognitive-services/speaker-recognition/},
  2022.

\bibitem{advertising}
{Popular Mechanics}, ``{Hundreds of Apps Can Eavesdrop Through Phone
  Microphones to Target Ads},''
  \url{https://www.popularmechanics.com/technology/security/a14533262/alphonso-audio-ad-targeting/},
  2018.

\bibitem{voicelone}
{BBC News}, ``Voice cloning of growing interest to actors and cybercriminals,''
  \url{https://www.bbc.com/news/business-57761873}, 2021.

\bibitem{jin2009v}
Q.~Jin, A.~R. Toth, T.~Schultz, and A.~W. Black, ``Voice convergin: Speaker
  de-identification by voice transformation,'' in \emph{Proceedings of {IEEE}
  {ICASSP}}, Taipei, Taiwan, 2009, pp. 3909--3912.

\bibitem{abou2015a}
M.~Abou{-}Zleikha, Z.~Tan, M.~G. Christensen, and S.~H. Jensen, ``A
  discriminative approach for speaker selection in speaker de-identification
  systems,'' in \emph{Proceedings of {IEEE} {EUSIPCO}}, Nice, France, 2015, pp.
  2102--2106.

\bibitem{magarinos2016p}
C.~Magari{\~{n}}os, P.~Lopez{-}Otero, L.~D. Fern{\'{a}}ndez, E.~R. Banga,
  C.~Garc{\'{\i}}a{-}Mateo, and D.~Erro, ``Piecewise linear definition of
  transformation functions for speaker de-identification,'' in
  \emph{Proceedings of IEEE {SPLINE}}, Aalborg, Denmark, 2016, pp. 1--5.

\bibitem{vaidya2019y}
T.~Vaidya and M.~Sherr, ``You talk too much: Limiting privacy exposure via
  voice input,'' in \emph{Proceedings of {IEEE} {S\&P} Workshops}, San
  Francisco, CA, USA, 2019, pp. 84--91.

\bibitem{qian2018h}
J.~Qian, H.~Du, J.~Hou, L.~Chen, T.~Jung, and X.~Li, ``Hidebehind: Enjoy voice
  input with voiceprint unclonability and anonymity,'' in \emph{Proceedings of
  {ACM} SenSys}, Shenzhen, China, 2018, pp. 82--94.

\bibitem{zhang2020e}
G.~Zhang, S.~Ni, and P.~Zhao, ``Enhancing privacy preservation in speech data
  publishing,'' \emph{{IEEE} Internet Things J.}, vol.~7, no.~8, pp.
  7357--7367, 2020.

\bibitem{qian2018t}
J.~Qian, F.~Han, J.~Hou, C.~Zhang, Y.~Wang, and X.~Li, ``Towards
  privacy-preserving speech data publishing,'' in \emph{Proceedings of {IEEE}
  {INFOCOM}}, Honolulu, HI, USA, 2018, pp. 1079--1087.

\bibitem{abs1905}
F.~Fang, X.~Wang, J.~Yamagishi, I.~Echizen, M.~Todisco, N.~W.~D. Evans, and
  J.~Bonastre, ``Speaker anonymization using x-vector and neural waveform
  models,'' \emph{CoRR}, vol. abs/1905.13561, 2019.

\bibitem{han2020v}
Y.~Han, S.~Li, Y.~Cao, Q.~Ma, and M.~Yoshikawa, ``Voice-indistinguishability:
  Protecting voiceprint in privacy-preserving speech data release,'' in
  \emph{Proceedings of {IEEE} {ICME}}, London, UK, 2020, pp. 1--6.

\bibitem{mohammadi2017a}
S.~H. Mohammadi and A.~Kain, ``An overview of voice conversion systems,''
  \emph{Speech Commun.}, vol.~88, pp. 65--82, 2017.

\bibitem{srivastava2020d}
B.~M.~L. Srivastava, N.~A. Tomashenko, X.~Wang, E.~Vincent, J.~Yamagishi,
  M.~Maouche, A.~Bellet, and M.~Tommasi, ``Design choices for x-vector based
  speaker anonymization,'' in \emph{Proceedings of {ISCA} Interspeech}, Virtual
  Event, Shanghai, China, 2020, pp. 1713--1717.

\bibitem{turner2022g}
H.~Turner, G.~Lovisotto, and I.~Martinovic, ``Generating identities with
  mixture models for speaker anonymization,'' \emph{Comput. Speech Lang.},
  vol.~72, p. 101318, 2022.

\bibitem{justin2015s}
T.~Justin, V.~Struc, S.~Dobrisek, B.~Vesnicer, I.~Ipsic, and F.~Mihelic,
  ``Speaker de-identification using diphone recognition and speech synthesis,''
  in \emph{Proceedings of {IEEE} {FG}}, Ljubljana, Slovenia, 2015, pp. 1--7.

\bibitem{ahmed2020p}
S.~Ahmed, A.~R. Chowdhury, K.~Fawaz, and P.~Ramanathan, ``Preech: {A} system
  for privacy-preserving speech transcription,'' in \emph{Proceedings of
  {USENIX} Security}, Virtual Event, 2020, pp. 2703--2720.

\bibitem{srivastava2020e}
B.~M.~L. Srivastava, N.~Vauquier, M.~Sahidullah, A.~Bellet, M.~Tommasi, and
  E.~Vincent, ``Evaluating voice conversion-based privacy protection against
  informed attackers,'' in \emph{Proceedings of {IEEE} {ICASSP}}, Barcelona,
  Spain, 2020, pp. 2802--2806.

\bibitem{goodfellow2014e}
I.~J. Goodfellow, J.~Shlens, and C.~Szegedy, ``Explaining and harnessing
  adversarial examples,'' in \emph{Proceedings of ICLR}, San Diego, CA, USA,
  2015.

\bibitem{madry2018t}
A.~Madry, A.~Makelov, L.~Schmidt, D.~Tsipras, and A.~Vladu, ``Towards deep
  learning models resistant to adversarial attacks,'' in \emph{Proceedings of
  ICLR}, Vancouver, BC, Canada, 2018.

\bibitem{carlini2017t}
N.~Carlini and D.~A. Wagner, ``Towards evaluating the robustness of neural
  networks,'' in \emph{Proceedings of {IEEE} {S\&P}}, San Jose, CA, USA, 2017,
  pp. 39--57.

\bibitem{Yuan2018c}
X.~Yuan, Y.~Chen, Y.~Zhao, Y.~Long, X.~Liu, K.~Chen, S.~Zhang, H.~Huang,
  X.~Wang, and C.~A. Gunter, ``Commandersong: {A} systematic approach for
  practical adversarial voice recognition,'' in \emph{Proceedings of {USENIX}
  Security}, Baltimore, MD, USA, 2018, pp. 49--64.

\bibitem{li2020a}
Z.~Li, Y.~Wu, J.~Liu, Y.~Chen, and B.~Yuan, ``Advpulse: Universal,
  synchronization-free, and targeted audio adversarial attacks via subsecond
  perturbations,'' in \emph{Proceedings of {ACM} CCS}, Virtual Event, USA,
  2020, pp. 1121--1134.

\bibitem{chen2021w}
G.~Chen, S.~Chen, L.~Fan, X.~Du, Z.~Zhao, F.~Song, and Y.~Liu, ``Who is real
  bob? adversarial attacks on speaker recognition systems,'' in
  \emph{Proceedings of {IEEE} {S\&P}}, San Francisco, CA, USA, 2021, pp.
  694--711.

\bibitem{phoneytalker}
M.~Chen, L.~Lu, Z.~Ba, and K.~Ren, ``Phoneytalker: An out-of-the-box toolkit
  for adversarial example attack on speaker recognition,'' in \emph{Proceedings
  of {IEEE} {INFOCOM}}, London, United Kingdom, 2022, pp. 1419--1428.

\bibitem{qin2019a}
Y.~Qin, N.~Carlini, G.~Cottrell, I.~Goodfellow, and C.~Raffel, ``Imperceptible,
  robust, and targeted adversarial examples for automatic speech recognition,''
  in \emph{Proceedings of ICML}, vol.~97, Long Beach, California, 2019, pp.
  5231--5240.

\bibitem{Schonherr2019}
L.~Sch{\"{o}}nherr, K.~Kohls, S.~Zeiler, T.~Holz, and D.~Kolossa, ``Adversarial
  attacks against automatic speech recognition systems via psychoacoustic
  hiding,'' in \emph{Proceedings of {NDSS}}, San Diego, California, USA, 2019.

\bibitem{Abdullah2019}
H.~Abdullah, W.~Garcia, C.~Peeters, P.~Traynor, K.~R.~B. Butler, and J.~Wilson,
  ``Practical hidden voice attacks against speech and speaker recognition
  systems,'' in \emph{Proceedings of {NDSS}}, San Diego, California, USA, 2019.

\bibitem{Wang2020}
Q.~Wang, P.~Guo, and L.~Xie, ``Inaudible adversarial perturbations for targeted
  attack in speaker recognition,'' in \emph{Proceedings of {ISCA} Interspeech},
  Virtual Event, 2020, pp. 4228--4232.

\bibitem{hussainNDMK21}
S.~Hussain, P.~Neekhara, S.~Dubnov, J.~J. McAuley, and F.~Koushanfar,
  ``Waveguard: Understanding and mitigating audio adversarial examples,'' in
  \emph{Proceedings of {USENIX} Security}, 2021, pp. 2273--2290.

\bibitem{Eisenhofer2021}
T.~Eisenhofer, L.~Sch{\"{o}}nherr, J.~Frank, L.~Speckemeier, D.~Kolossa, and
  T.~Holz, ``Dompteur: Taming audio adversarial examples,'' in
  \emph{Proceedings of {USENIX} Security}, 2021, pp. 2309--2326.

\bibitem{voiceprint}
{Alibaba Cloud}, ``Voiceprint recognition system — not just a powerful
  authentication tool,''
  \url{https://alibaba-cloud.medium.com/voiceprint-recognition-system-not-just-a-powerful-authentication-tool-6b3702b5c5a},
  2017.

\bibitem{shortcut}
Apple, ``{Shortcuts User Guide},''
  \url{https://support.apple.com/en-hk/guide/shortcuts/welcome/ios}, 2022.

\bibitem{anywhere}
Absinthe, ``{Anywhere},''
  \url{https://play.google.com/store/apps/details?id=com.absinthe.anywhere_},
  2021.

\bibitem{snyder2018x}
D.~Snyder, D.~Garcia-Romero, G.~Sell, D.~Povey, and S.~Khudanpur, ``X-vectors:
  Robust dnn embeddings for speaker recognition,'' in \emph{Proceedings of IEEE
  ICASSP}, Calgary, AB, Canada, 2018, pp. 5329--5333.

\bibitem{speechbrain}
M.~Ravanelli, T.~Parcollet, P.~Plantinga, A.~Rouhe, S.~Cornell, L.~Lugosch,
  C.~Subakan, N.~Dawalatabad, A.~Heba, J.~Zhong, J.~Chou, S.~Yeh, S.~Fu,
  C.~Liao, E.~Rastorgueva, F.~Grondin, W.~Aris, H.~Na, Y.~Gao, R.~D. Mori, and
  Y.~Bengio, ``Speechbrain: {A} general-purpose speech toolkit,'' \emph{CoRR},
  vol. abs/2106.04624, 2021.

\bibitem{panayotov2015l}
V.~Panayotov, G.~Chen, D.~Povey, and S.~Khudanpur, ``Librispeech: An {ASR}
  corpus based on public domain audio books,'' in \emph{Proceedings of IEEE
  ICASSP}, South Brisbane, Queensland, Australia, 2015, pp. 5206--5210.

\bibitem{schultzJ2003l}
M.~Schultz and T.~Joachims, ``Learning a distance metric from relative
  comparisons,'' in \emph{Proceedings of NIPS}, Vancouver and Whistler, British
  Columbia, Canada, 2003, pp. 41--48.

\bibitem{dehak2010front}
N.~Dehak, P.~J. Kenny, R.~Dehak, P.~Dumouchel, and P.~Ouellet, ``Front-end
  factor analysis for speaker verification,'' \emph{IEEE Transactions on Audio,
  Speech, and Language Processing}, vol.~19, no.~4, pp. 788--798, 2010.

\bibitem{desplanques2020e}
B.~Desplanques, J.~Thienpondt, and K.~Demuynck, ``{ECAPA-TDNN:} emphasized
  channel attention, propagation and aggregation in {TDNN} based speaker
  verification,'' in \emph{Proceedings of ISCA Interspeech}, Shanghai, China,
  2020, pp. 3830--3834.

\bibitem{fayek2016s}
H.~M. Fayek, ``Speech processing for machine learning: Filter banks,
  mel-frequency cepstral coefficients (mfccs) and what's in-between,''
  \url{https://haythamfayek.com/2016/04/21/speech-processing-for-machine-learning.html},
  2016.

\bibitem{kingmaW2013a}
D.~P. Kingma and M.~Welling, ``Auto-encoding variational bayes,'' in
  \emph{Proceedings of ICLR}, Banff, AB, Canada, 2014.

\bibitem{hsu2017l}
W.~Hsu, Y.~Zhang, and J.~R. Glass, ``Learning latent representations for speech
  generation and transformation,'' in \emph{Proceedings of ISCA Interspeech},
  Stockholm, Sweden, 2017, pp. 1273--1277.

\bibitem{variani2014d}
E.~Variani, X.~Lei, E.~McDermott, I.~L. Moreno, and J.~Gonzalez-Dominguez,
  ``Deep neural networks for small footprint text-dependent speaker
  verification,'' in \emph{Proceedings of IEEE ICASSP}, Florence, Italy, 2014,
  pp. 4052--4056.

\bibitem{li2017d}
L.~Li, Y.~Chen, Y.~Shi, Z.~Tang, and D.~Wang, ``Deep speaker feature learning
  for text-independent speaker verification,'' in \emph{Proceedings of ISCA
  Interspeech}, Stockholm, Sweden, 2017, pp. 1542--1546.

\bibitem{hugginface}
{Hugging Face}, ``Open-source pretrained models.''
  \url{https://huggingface.co/models}, 2022.

\bibitem{cmu_speech}
C.~S. Group, ``Statistical parametirc sythesis and voice conversion
  techniques,'' \url{http://festvox.org/11752/slides/lecture11a.pdf}, 2012.

\bibitem{chouL2019}
J.~Chou and H.~Lee, ``One-shot voice conversion by separating speaker and
  content representations with instance normalization,'' in \emph{Proceedings
  of ISCA Interspeech}, Graz, Austria, 2019, pp. 664--668.

\bibitem{mohan2020e}
B.~M.~L. Srivastava, N.~Vauquier, M.~Sahidullah, A.~Bellet, M.~Tommasi, and
  E.~Vincent, ``Evaluating voice conversion-based privacy protection against
  informed attackers,'' in \emph{Proceedings of IEEE ICASSP}, Barcelona, Spain,
  2020, pp. 2802--2806.

\bibitem{yangLCS19}
Z.~Yang, B.~Li, P.~Chen, and D.~Song, ``Characterizing audio adversarial
  examples using temporal dependency,'' in \emph{Proceedings of {ICLR}}, New
  Orleans, LA, USA, 2019.

\bibitem{siri}
{Apple}, ``{Apple Siri},''
  \url{https://machinelearning.apple.com/research/personalized-hey-siri}, 2022.

\bibitem{pobar2014o}
M.~Pobar and I.~Ipsic, ``Online speaker de-identification using voice
  transformation,'' in \emph{Proceedings of IEEE {MIPRO}}, Opatija, Croatia,
  2014, pp. 1264--1267.

\bibitem{magarinos2017r}
C.~Magari{\~{n}}os, P.~Lopez{-}Otero, L.~D. Fern{\'{a}}ndez, E.~R. Banga,
  D.~Erro, and C.~Garc{\'{\i}}a{-}Mateo, ``Reversible speaker de-identification
  using pre-trained transformation functions,'' \emph{Comput. Speech Lang.},
  vol.~46, pp. 36--52, 2017.

\bibitem{eide1996a}
E.~Eide and H.~Gish, ``A parametric approach to vocal tract length
  normalization,'' in \emph{Proceedings of IEEE {ICASSP}}, Atlanta, Georgia,
  USA, 1996, pp. 346--348.

\bibitem{srivastava2019p}
B.~M.~L. Srivastava, A.~Bellet, M.~Tommasi, and E.~Vincent,
  ``Privacy-preserving adversarial representation learning in {ASR:} reality or
  illusion?'' in \emph{Proceedings of {ISCA} Interspeech}, Graz, Austria, 2019,
  pp. 3700--3704.

\bibitem{abs2011}
F.~M. Espinoza{-}Cuadros, J.~M. Perero{-}Codosero,
  J.~Ant{\'{o}}n{-}Mart{\'{\i}}n, and L.~A.~H. G{\'{o}}mez, ``Speaker
  de-identification system using autoencoders and adversarial training,''
  \emph{CoRR}, vol. abs/2011.04696, 2020.

\end{thebibliography}
    \begin{appendices}
	\section{MOS Test Descriptions}
\label{appendix-mos}
In the comparing trial, each volunteer needs to listen to the original and de-identified voices and compare their similarity in terms of voiceprint, speech text and audio quality. Their opinions are recorded as 5-level MOS ratings, whose detailed description is shown in Table~\ref{tab:mos-com}.

\begin{table}[hb]
    \centering
    \caption{5 MOS levels in the comparing trial.}
    \label{tab:mos-com}
    \renewcommand{\arraystretch}{1.25}
    \begin{tabularx}{\linewidth}{m{0.1\linewidth}<{\centering}|m{0.8\linewidth}}
    \hline
        \textbf{Level} & \textbf{Description} \\
    \hline
        5 & The voiceprint/text/quality of the test audio is the same as that of the reference audio and there is no difference. \\
    \hline
        4 & The voiceprint/text/quality of the test audio is relatively close to that of the reference audio and there is no obvious difference. \\
    \hline
        3 & The voiceprint/text/quality of the test audio is roughly similar to that of the reference audio but there is slight difference. \\
    \hline
        2 & The voiceprint/text/quality of the test audio differs to that of the reference audio and there is perceivable difference. \\
    \hline
        1 & The voiceprint/text/quality of the test audio is totally different from that of the reference audio and there is quite obvious difference. \\
    \hline
    \end{tabularx}
\end{table}

In the distinguishing trial, each volunteer need to listen to the original and de-identified voices in a random order and determines whether they are original or not. If the volunteers regard it as not original, we further require them to select the reason for their judgement from several options or provide their own evidence, whose detailed description is shown in Table~\ref{tab:mos-dis}.

\begin{table}[hb]
    \centering
    \caption{Option and reasons in the distinguishing trial.}
    \label{tab:mos-dis}
    \renewcommand{\arraystretch}{1.25}
    \begin{tabularx}{\linewidth}{m{0.3\linewidth}<{\centering}|m{0.6\linewidth}}
    \hline
        \textbf{Option} & \textbf{Description} \\
    \hline
        Yes & The test audio is the original voice. \\
    \hline
        No & The test audio is not the original voice. \\
    \hline\hline
        \textbf{Reason} & \textbf{Description} \\
    \hline
        Unnatural Voiceprint & The voiceprint of the test audio sounds synthetic but not real human voice. \\
    \hline
        Illegible Text & The text of the test audio is corrupted and hard to recognize. \\
    \hline
        Distorted Quality & The quality of the test audio is distorted. \\
    \hline
        Obvious Reverb & There are reverb echos in the test audio. \\
    \hline
        Other Reason & Other reasons from volunteers. \\
    \hline
    \end{tabularx}
\end{table}

	\end{appendices}
    \balance

\end{document}